\documentclass[12pt,groupaddress,showpacs,floatfix]{revtex4}

\usepackage{graphicx}% Include figure files
\usepackage{dcolumn}% Align table columns on decimal point
\usepackage{bm}% bold math
\usepackage{amsmath}
\usepackage{amssymb}
\usepackage{times}

\usepackage{amsmath,amssymb,enumerate,bbm}

\begin{document}

\title{Approximating the satisfiability transition by suppressing fluctuations.}

\author{S. Knysh}
\email{knysh@email.arc.nasa.gov}

\author{V. N. Smelyanskiy}
\email{Vadim.N.Smelyanskiy@nasa.gov}

\author{R.D. Morris}
\email{rdm@email.arc.nasa.gov}

\affiliation{NASA Ames Research Center, MS 269-3, Moffett Field, CA 94035-1000}

\date{\today}

\begin{abstract}
Using methods and ideas
from statistical mechanics and random graph theory, we propose a
simple method for obtaining rigorous upper bounds for the
satisfiability transition in random boolean expressions composed
of $N$ variables and $M$ clauses with $K$ variables per clause.
 The method is based on the identification of the
core -- a subexpression (subgraph) that has the same
satisfiability properties as the original expression. We formulate
self-consistency equations that determine the macroscopic
parameters of the core and compute an improved annealing bound for
the satisfiability threshold, $\alpha_c=M/N$.  We illustrate the
method for three sample problems: $K$-XOR-SAT, $K$-SAT and
positive $1$-in-$K$-SAT.
\end{abstract}

 \maketitle

\section{Introduction}

Over the past decade the statistical properties of combinatorial problems has
attracted increasingly greater attention from both the computer science and
physics communities \cite{ksat1,ksat2,ksat3}.
Most computationally difficult problems encountered in practice belong to the class of
NP-complete problems.
There is a one-to-one correspondence between
these problems and spin glass models \cite{fu}. Unlike problems with regular structure, many
combinatorial optimization problems are formulated on random graphs and hypergraphs.
The long-standing problem in the computer science
community is \lq\lq P vs. NP'', that is, can NP-complete problems
 be solved in polynomial time, or they are inherently intractable \cite{karp}? Although
the problem is extremely important, it is also deeply theoretical as it
concentrates on worst-case scenarios. From the viewpoint of practitioners,
efficient algorithms have to be designed with real-world problems in mind.
Appropriate test cases can be prepared for comparing the performance of
different algorithms. However, this approach does not allow the design
of algorithms with typical performance in mind, only their comparison.

One can argue that the purely theoretical study of algorithms was
somewhat impeded by exploding speed of computers that encouraged
experimentation. This state of affairs may be challenged by the emerging
paradigm of quantum computing. Until a working prototype of a quantum computer
is built, it exists only on paper. Classical simulations of quantum computers
can be done only for very small problems, due to speed and memory
requirements.  Since these reqirements grow exponentially with the size of the
problem, they could be used only in proof-of-concept scenarios. While quantum
computation was shown to be efficient for some classically intractable
problems (the most notable example being Shor's algorithm \cite{shor}), whether they provide
an advantage for NP-complete problems is unresolved. Therefore, designing algorithms
with typical complexity in mind for quantum computer may be desirable.
Whether the newly proposed quantum adiabatic algorithm is efficient in tackling
NP-complete problems is an area of active research \cite{farhi,vazirani,hogg}.

The statistics of real-world examples is largely unknown. As a first
approximation one can assume that the problems can be chosen completely at
random. The underlying belief is that if an algorithm is efficient for a
uniform ensemble of randomly chosen problem instances, it will solve real-world
examples fairly efficiently as well.
The performance for random problems is a truly unbiased benchmark to compare
different algorithms. The same explosion in computational speed responsible
for diminished reliance on theoretical study has also reignited  interest in this type of study.

Many problems of interest are written as a boolean expression (a formula) --
a set of $N$ variables and $M$ constraints, all which we aim to satisfy.
Each constraint is a clause involving $K$ variables and it determines which combinations of variables are
permitted. The types of constraints differ from problem to problem, but for
great many the following picture persists:
for small $\alpha = M / N$ the
problem is almost always (that is, with probability $1$ in the limit $N
\rightarrow \infty$) satisfiable, while at $\alpha = \alpha_c$ an abrupt change
occurs, and for all $\alpha > \alpha_c$ the problem is almost always
unsatisfiable \cite{ksat1,ksat2,ksat3}. An even more interesting phenomenon occurs for the typical running
time of the algorithm: the time it takes to solve a problem is usually
polynomial for $\alpha < \alpha_d < \alpha_c$, and exponential for $\alpha >
\alpha_d$, where $\alpha_d$ is algorithm-dependent. However, independent of
the algorithm used, the complexity peaks at $\alpha = \alpha_c$, where the
probability that the formula is satisfiable is approximately $1 / 2$.

Random satisfiability problems grabbed the attention of the statistical
physics community, since the phenomenon in question is a phase transition; the
study of this phase transition may improve the understanding of the physics of random
materials such as glasses. This is in addition to any statistical properties
of the solutions --  properties that can be used for the design of efficient
classical or quantum algorithms.

The quest for exact values of $\alpha_c$ or $\alpha_d$ has  so far been
elusive. The best results for a particular problem -- K-SAT -- were obtained
using the so-called one-step RSB approximation and are in excellent agreement with
experiment \cite{ksat1rsb}. However, the method has drawn criticism because the method itself
is not well-understood, lacks a rigorous foundation, and the result depends on
extensive numerical computations. On the upside, rigorous bounds have been
obtained for $K$-XOR-SAT (note, however that it can be solved in polynomial
time). On the mathematical side, a series of results on rigorous lower \cite{lower} and
upper \cite{upper} bounds on $\alpha_c$ appeared recently. Typically lower bounds rely
on an explicit algorithm and upper bounds rely on the counting of solutions. The
trivial upper bound is obtained using the annealing approximation. All
improvements over the annealing approximation employ the fact that at the
satisfiability transition the number of solutions jumps from the exponentially
large number $2^{\alpha N}$ to $0$. The method we propose in this paper does
not deviate from this strategy. For any random formula we identify a
subformula that possesses identical satisfiability properties, but has suppressed
fluctuations. That is, if the formula is satisfiable, the subformula is also
satisfiable but has a significantly smaller number of solutions. By performing
the disorder average of the number of solutions of the subformula (rather than
formula, as in the annealing approximation) the point where the average goes to
zero determines the upper bound on the true transition point.

The advantages of the method described here are that it is rigorous (it does not
rely on any hypotheses, although we supply proofs only when they are not
immediately intuitive; it is straightforward to rederive all the results with
complete mathematical rigor) and that the method is applicable to various
types of random satisfiability problems. We choose to describe $K$-XOR-SAT as well
as the NP-complete problems $K$-SAT and positive $1$-in-$K$-SAT. Each problem adds its own \lq\lq touch''
to the formalism. For the case $K$-XOR-SAT -- a polynomial problem -- the upper
bound is exact \cite{kxorsat}, while the upper bound for $K$-SAT grossly overestimates
the transition. This could be related to the fact that $K$-SAT is very difficult
for classical algorithms.
In all cases we take a two step approach. In the first step
we compute the parameters of the subformula -- the core. In the second step we
compute the annealing approximation for the number of solutions of the
subformula. The size of the core also exhibits a phase transition and has been
studied for a range of problems \cite{pittel}. Our method provides a much simpler way to
derive those results.

The paper is organized as follows. We describe $K$-XOR-SAT, $K$-SAT and
positive $1$-in-$K$-SAT in sections \ref{sec:KXORSAT} through \ref{sec:1inKSAT};
section \ref{sec:Simulations} is devoted to
numerical simulations for  the positive $1$-in-$3$-SAT problem; section \ref{sec:Summary}
is a summary.

\section{\label{sec:KXORSAT}$K$-XOR-SAT}

In this model the instance of the problem consists of $N$ variables and $M$
clauses, each clause involving $K$ variables. Each variable can take values
$0$ or $1$. The ensemble we consider (random hypergraph) is that of independent
clauses with variables in each clause drawn uniformly at random out of the set
of $N$ variables. To each clause we also attribute a number $0$ or $1$, each
with probability of $1 / 2$, and posit that the clause is satisfied if the
exclusive-or (XOR) of the $K$ variables in the clause equals that number. The entire
formula is said to be satisfied if all of its clauses are satisfied

The probability that such random formula is satisfied, in the limit $N
\rightarrow \infty$, exhibits a sharp jump from $1$ to $0$ at some critical
ratio of clauses to variables $\alpha_c = M / N$. We attempt to estimate this
satisfiability threshold. The simplest approximation (in fact an upper bound)
uses the first moment method (known as the annealing approximation in the physics
community). One can compute the disorder-averaged number of solutions. The
point where the expectation value of the number of solutions becomes smaller
than $1$ corresponds to a formula that is unsatisfiable; therefore this serves
as an upper bound on the location of the transition. In essence we have
approximated $\mathbbm{P}( \textrm{sat} ) \equiv \mathbbm{P}[\mathcal{N}
\geqslant 1 ] =\mathbbm{E}[ \theta (\mathcal{N}- 1 ) ]$ by
$\mathbbm{E}[\mathcal{N}]$, where $\mathcal{N}$ denotes the number of
solutions (an integer). In the physics community the annealing approximation for the
entropy is regarded as the replacement of the correct quantity $\mathbbm{E}[ \ln
\mathcal{N}]$ by the incorrect expression $\ln \mathbbm{E}[\mathcal{N}]$.

Computing the point where the annealed entropy becomes zero is trivial. For
each clause, the probability that the clause is satisfiable is independent of the
assignment of variables and equals $1 / 2$. Therefore the expected number of
solutions is
\begin{equation}
\mathbbm{E}[\mathcal{N}] = 2^N 2^{- M},
\end{equation}
and the corresponding entropy $S_{\textrm{ann}} = N \ln 2 - M \ln 2$ becomes
negative above $\alpha_u = 1$ (the subscript indicates that this is the upper
bound).

\subsection{Concept of a {\textit{core}}}

A major drawback of the annealing approximation is in that it fails to account for
finite entropy at the satisfiability transition. (By accident, for this
particular problem, the annealed expression for the entropy on the satisfiable
side of the transition is exact). It can be argued that at any finite
connectivity a random graph possesses a large ($O ( N )$) number of
variables that do not appear in any clauses, thus making a contribution to the
entropy which we fail to take into account. Furthermore, there are clauses
that involve variables, the variables not being in any other clauses, as well as small clusters
of such clauses. The annealing bound would be significantly improved if it
were possible to separate these {\textit{irrelevant}} contributions to the
entropy.

In a paper devoted to the finite-size effects of the satisfiability transition, a
concept of {\textit{irrelevant}} clauses was put forward. Given a random formula
one can always easily identify clauses that can be trivially satisfied. The
paper did not specify the procedure for finding such clauses, only that their
number is extensive ($O ( N )$). One example is isolated clauses, since
variables can always be set so as to satisfy the clause. The presence of such
extensive clauses is responsible for the lower bound of $2$ of the finite-size
scaling exponent $\nu$, or, in other words, that the disorder is relevant to
the phase transition.

One can try to advance the most general definition of irrelevant clause based
on local properties alone. In fact this has been done for $K$-XOR-SAT \cite{kxorsat}. In
essence we repeat the derivation in a slightly simplified form, but will
generalize it to other problems later on. For $K$-XOR-SAT we identify
variables that appear in no clauses and delete those variables. Further, we
identify variables that appear in only one clause. Such variables can be set
to $0$ or $1$ (after other variables have been assigned) so that the clause
becomes satisfiable. Hence the satisfiability of the entire formula will be
unaffected if the variable and the corresponding clause are deleted. This
process (known as trimming algorithm, illustrated below, in Fig. \ref{fig:trim1}) can be continued
until  we either end up with an empty graph (which would imply that the
formula is satisfiable) or a {\textit{core}} -- the formula in which all
variables appear in at least two clauses. One can compute the annealed entropy
on the core and use the point at which the entropy becomes zero as the improved
upper bound $\alpha_u'$.

\begin{figure}[!ht]
\includegraphics[width=2.5in]{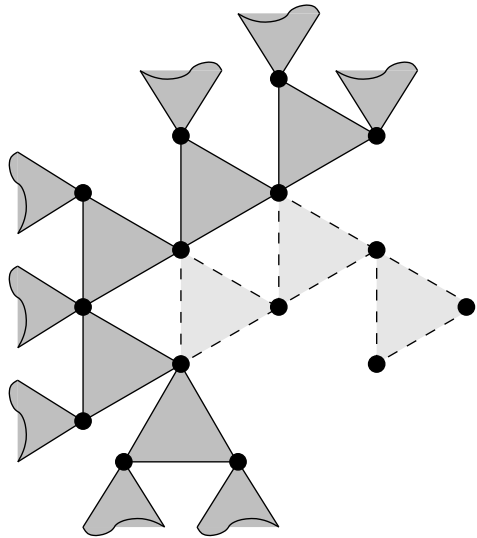}
\caption{\label{fig:trim1}Example of trimming algorithm for 3-XOR-SAT. Variables are
represented graphically as vertices and clauses are represented as triangles. Incomplete
triangles represent connections to the remainder of the graph (not shown). Lightly shaded clauses
are removed by the trimming algorithm.}
\end{figure}

We examine the structure of the remaining core. First, observe that the
remaining core does not depend on the order in which the variables and clauses
are removed. In fact the remaining core is the unique maximal subformula of
the original formula with the property that every variable appears in at least
two clauses. The original formula is the core plus all deleted clauses and
variables. Assume that the core has $N'$ variables and $M'$ edges (implying $N
- N'$ variables and $M - M'$ clauses were deleted). Correspondingly, all
original graphs can be divided into distinct groups based on values of $N'$,
$M'$. Suppose we keep $N'$ and $M'$ fixed. Observe that to every realization
of the core there corresponds an equal number of possible realizations of
deleted clauses, and, as a consequence, an equal number of possible
realizations of the original graph (in the group labeled by $N'$, $M'$).
Therefore, for any fixed $N'$, $M'$ all possible realization of the core are
equiprobable -- a fact we employ to perform disorder averages.

Notice that all possible realizations of the core are equiprobable only for
fixed $N'$, $M'$. The values of $N'$ and $M'$ themselves fluctuate. However,
the fluctuations in $N'$ and $M'$ are on the order of $O ( \sqrt{N} )$ while
their respective values are $O ( N )$. Since we expect the threshold to be
sharp as a function of $M' / N'$, we need not concern ourselves with these
fluctuations. Therefore we concentrate on finding the most likely values of
$N' / N$ and $M' / N$. One approach is to work with a set of $\{ c_k \}$ -- a
fraction of vertices that appear in $k$ clauses. One can describe an algorithm
as a random process and study the changes in the average values of $\{ c_k \}$.
The discrete steps of the algorithm are approximated by continuous time $t$, and a
set of $\{ c_k ( t ) \}$ is replaced by its generating function $c ( t, x ) =
\sum_k c_k ( t ) x^k$. The problem is then reduced to solving the resulting
PDE. This is the approach taken in \cite{kxorsat}. Slightly differing variants
of this method were also employed in \cite{pittel,trim1,trim2}.
We instead opt for an approach that does not take dynamics into considerations.
The approach is inspired in part by work analyzing the matching problem \cite{karpsipser}.

In essence, we seek the disordered average of $N' / N$. This is precisely the
probability $p$ that a randomly chosen variable belongs to the core. We can
also fix a specific variable (say, variable $x_0$) and perform a disorder
average of a function that yields $1$ if that vertex belongs to the core or
$0$ if it does not. For every formula $\mathcal{F}$ we can introduce the set
$\mathcal{C}$ of variables that belong to the core. Obviously $|\mathcal{C}|
\equiv N' = pN$. Now, introduce an extension of $\mathcal{C}$, which we denote
as $\mathcal{C}'$, defined as the minimal set that satisfies the following
requirements
\begin{enumerate}
  \item $\mathcal{C} \subseteq \mathcal{C}'$.

  \item If $K - 1$ variables in some clause belong to $\mathcal{C}'$, then the
  remaining variable must also belong to $\mathcal{C}'$.
\end{enumerate}
It is straightforward to see that set $\mathcal{C}'$ so defined is unique. Let
$|\mathcal{C}' | = qN$, where $q$ can be interpreted as the probability that a
random vertex belongs to $\mathcal{C}'$.

Let us turn to the original random graph. The number of clauses in which
the variable $x_0$ appears is a random variable distributed according to a Poisson
distribution with parameter $K \alpha$. In performing disorder averages we can
first average over all possible disorders with fixed values of clauses $k$
first, and average over $k$ with weight $\mathrm{e}^{- K \alpha} ( K \alpha )^k /
k!$ as the last step. Further, observe that those $k$ clauses are independent.
Let $\mathcal{F}'$ denote a formula that is obtained by removing the variable
$x_0$ and the clauses in which it appears. Let $q'$ denote the parameter $q$
associated with $\mathcal{F}'$. Suppose that for some clause in which $x_0$
appears, all the other $K - 1$ variables belong to $\mathcal{C}' [\mathcal{F}' ]$.
Then $x_0$ must belong to $\mathcal{C}[\mathcal{F}]$. The probability that for
some clause $K - 1$ variables other than $x_0$ belong to $C' [\mathcal{F}' ]$
is $( q' )^{K - 1}$. The number of such clauses is, hence, also Poisson, but
with parameter $K \alpha ( q' )^{K - 1}$. The probability $q$ that $x_0$
belongs to $\mathcal{C}[\mathcal{F}]$ is therefore
\begin{equation}
q = \sum_{k = 1}^{\infty} \mathrm{e}^{- K \alpha ( q' )^{K - 1}} \frac{\left( K
   \alpha ( q' )^{K - 1} \right)^k}{k!} = 1 - \mathrm{e}^{- K \alpha ( q' )^{K -
   1}} .
\end{equation}
Now observe that $\mathcal{F}'$ is essentially a random formula with $N - 1$
variables and the same (to within $O ( 1 / N )$) ratio of clauses to variables.
Therefore in the limit $N \rightarrow \infty$ which we are ultimately
interested in, there should be no difference in statistical properties, and
hence $q = q'$. This leads to self-consistency equation
\begin{equation}
q = 1 - \mathrm{e}^{- K \alpha q^{K - 1}} . \label{eq:KXOR-q}
\end{equation}
Note that $q = 0$ is always a solution to this equation. Since the core is
defined as the largest possible subformula with certain properties, and the
size of the core is directly related to $q$, we must adopt a convention that
the largest possible solution to (\ref{eq:KXOR-q}) is always chosen. Below a certain threshold
only $q = 0$ is a solution, whereas above the threshold, another $q > 0$
solution appears.

We now turn to the original goal of finding $N'$. If at least two clauses
which include $x_0$ have the property that $K - 1$ other variables are in $\mathcal{C}'
[\mathcal{F}' ]$, then the variable $x_0$ as well as the aforementioned variables
are in $\mathcal{C}[\mathcal{F}]$. Hence, we can write
\begin{equation}
   p = \sum_{k = 2}^{\infty} \mathrm{e}^{- K \alpha q^{K - 1}} \frac{\left( K
   \alpha q^{K - 1} \right)^k}{k!} = 1 - \left( 1 + K \alpha q^{K - 1} \right)
   \mathrm{e}^{- K \alpha q^{K - 1}} .
\end{equation}
To compute $M'$ we examine the average degree (number of clauses in which it
appears) of the randomly chosen vertex in the core. The latter should equal
$KM' / N'$. If vertex $x_0$ is in the core (with probability $p$), the number
of clauses which are in the core was shown above to be a random variable -- a
truncated (only $k \geqslant 2$ are allowed) Poisson distribution with
parameter $K \alpha q^{K - 1}$. Therefore
\begin{equation}
   KM' / N' = \sum_{k = 2}^{\infty} k \mathrm{e}^{- K \alpha q^{K - 1}}
   \frac{\left( K \alpha q^{K - 1} \right)^k}{k!} \Big/ \sum_{k = 2}^{\infty}
   \mathrm{e}^{- K \alpha q^{K - 1}} \frac{\left( K \alpha q^{K - 1}
   \right)^k}{k!} .
\end{equation}
Recognizing that the denominator is $p = N' / N$ we can rewrite
\begin{equation}
   M' / N = \frac{1}{K} \sum_{k = 2}^{\infty} k \mathrm{e}^{- K \alpha q^{K - 1}}
   \frac{\left( K \alpha q^{K - 1} \right)^k}{k!} = \alpha q^{K - 1} \left( 1
   - \mathrm{e}^{- K \alpha q^{K - 1}} \right) = \alpha q^K . \label{eq:KXOR-M}
\end{equation}

\subsection{Improved annealing bound}

As with the original annealing bound, we are aided by the fact that clauses
require that the exclusive-or of the variables be either $0$ or $1$ with
probability $1 / 2$. The probability that a clause is satisfied is independent
of the assignment of the variables, and the entropy is predicted to decrease to zero
when $M' / N' = 1$ or
\begin{equation}
\alpha q^K = q - K \alpha q^{K - 1} + K \alpha q^K
\end{equation}
Coupled with $1 - q = \mathrm{e}^{- K \alpha q^{K - 1}}$ this puts the upper bound
of critical threshold at $\alpha_u' \approx 0.918$.

It is a remarkable feature of $K$-XOR-SAT is that whenever it is satisfiable,
the number of solutions of $K$-XOR-SAT equals the number of solutions of
the corresponding \lq\lq ferromagnetic'' model, where we require that the exclusive-or of
the variables be precisely $0$ in all clauses. Note that for $K$-XOR-SAT, this is
not so; the ferromagnetic model always possesses at least one solution. The next
observation is that the disorder average of the square of the number of
solutions of $K$-XOR-SAT $\mathbbm{E}[\mathcal{N}^2 ]$ equals $2^{N' - M'}$
multiplied by $\mathbbm{E}[\mathcal{N}]$ as computed for the ferromagnetic model.
As long as the annealing bound for the ferromagnetic model equals that for
$K$-XOR-SAT we can be sure that the annealing bound is correct and we are in the
satisfiable phase. The point at which it ceases to be so is the lower bound on
the satisfiability transition $\alpha_l$.

Finding the annealed entropy for the ferromagnetic model on a complete graph is
trivial and amounts to finding a maximum of
\begin{equation}
   - N \Big[ \frac{1 + m}{2} \ln \frac{1 + m}{2} + \frac{1 - m}{2} \ln
   \frac{1 - m}{2} \Big] + M \ln \frac{1 + m^3}{2} .
\end{equation}
For as long as $m = 0$ is a global maximum of this expression, the annealed
entropies of the ferromagnetic and random models are equal. It ceases to be so at
$\alpha_l \approx 0.889$, which serves as a lower bound on satisfiability
transition. It is worthwhile to compute the annealed entropy on the core. That
task has been accomplished in \cite{kxorsat}. We rederive the results using a different
method which can be readily generalized to other problems.

The annealed entropy is simply the difference between $\ln \mathcal{N}_{s, J}$
and $\ln \mathcal{N}_J$, where $\mathcal{N}_J$ is the number of possible
disorders, and $\mathcal{N}_{s, J}$ counts the total number of disorder
configurations and variable assignments compatible with the disorder. For
simplicity, we decide to distinguish between disorders that differ only by
permutation of clauses and permutation of variables within clauses. Any double
counting in $\mathcal{N}_J$ due to this convention will be exactly canceled
by identical factor in $\mathcal{N}_{s, J}$. The advantages are especially
evident for the case of the original random graph. We can immediately obtain
$\mathcal{N}_J = N^{3 M}$. The expression is more complex when restricted so
that the degrees of all variables are at least $2$. We now investigate it closely.
We introduce a set $\{ c_{\boldsymbol{k}} \}$ where $\boldsymbol{k}$ is a vector
with $K$ components $\{ k_p \}$ that count the number of clauses in which some
variable appears in $p$-th position. The quantity $c_{\boldsymbol{k}}$ is the
fraction of variables described by vector $\boldsymbol{k}$. One trivial
constraint is that $\sum_{\boldsymbol{k}} c_{\boldsymbol{k}} = 1$. One can
represent disorders as an $M' \times K$ table of numbers from $1$ to $N'$. We
can divide the variables into various classes according to the value of
$\boldsymbol{k}$. The number of all possible permutations is the product of two
factors.
\begin{enumerate}
  \item $N' ! \big/ \prod_{\boldsymbol{k}} N_{\boldsymbol{k}}' !$ for the number of
  ways to arrange the variables into the various classes.

  \item $\prod_p \left[ M' ! \Big/ \prod_{\boldsymbol{k}} ( k_p !
  )^{N_{\boldsymbol{k}}'} \right]$ for the number of ways to rearrange the variables
  in the $M' \times K$ table.
\end{enumerate}
In general we ought to perform a sum over all possible values of
$N_{\boldsymbol{k}}'$. However, the sum is dominated by particular values of
$N_{\boldsymbol{k}}'$ that maximize the entropy ($N_{\boldsymbol{k}}' = N'
c_{\boldsymbol{k}}$)
\begin{eqnarray}
   S_J [ c_{\boldsymbol{k}} ] & = & - N' \sum_{\boldsymbol{k}} c_{\boldsymbol{k}} \ln
   c_{\boldsymbol{k}} + K ( M' \ln M' - M' ) - N' \sum_{\boldsymbol{k}}
   c_{\boldsymbol{k}}  \Big( \sum_p \ln k_p ! \Big) \nonumber \\
   & = & S_J^{( 1 )} [ c_{\boldsymbol{k}} ] + K ( M' \ln M' - M' ) .
\end{eqnarray}
Note that we have $K$ constraints on $c_{\boldsymbol{k}}$
\begin{equation}
\sum_{\boldsymbol{k}} k_p c_{\boldsymbol{k}} = M' / N', \label{eq:KXOR-constr}
\end{equation}
and that we require $c_{\boldsymbol{k}} = 0$ if $\sum_p k_p < 2$.

Maximizing $S_J^{( 1 )} [ c_{\boldsymbol{k}} ]$ is easiest if we work with its
dual transform. Let $\{ - \ln \mu_p \}$ be dual variables associated with
constraints (\ref{eq:KXOR-constr}). Instead of finding
\begin{equation}
   S_J^{( 1 )} [ N', M' ] = \max_{c_{\boldsymbol{k}}} \bigg\{ S_J^{( 0 )} [
   c_{\boldsymbol{k}} ] \, \bigg| \, \sum_{\boldsymbol{k}} k_p c_{\boldsymbol{k}} = M' / N'
   \bigg\}
\end{equation}
we compute
\begin{equation}
   \tilde{S}_J^{( 1 )} [ \{ \mu_p \} ] = \min_{M'_p} \bigg\{ - \sum_p M_p' \ln
   \mu_p - S_J^{(1)} [ N', \{ M_p' \} ] \bigg\},
\end{equation}
where $S_J^{( 1 )} [ N', \{ M_p' \} ]$ denotes $S_J^{( 1 )} [ c_{\boldsymbol{k}}
]$ maximized under the constraints $\sum_{\boldsymbol{k}} k_p c_{\boldsymbol{k}} =
M_p' / N'$. After simplifications
\begin{equation}
   \tilde{S}_J^{( 1 )} [ \{ \mu_p \} ] = \min_{\{ c_{\boldsymbol{k}} \}} \bigg\{
   - N' \sum_{\boldsymbol{k}} c_{\boldsymbol{k}} \sum_p k_p \ln \mu_p + N'
   \sum_{\boldsymbol{k}} c_{\boldsymbol{k}} \ln \Big[ c_{\boldsymbol{k}} \prod_p
   k_p ! \Big] \bigg\}
\end{equation}
Optimizing this with respect to $c_{\boldsymbol{k}}$ under the constraint
$\sum_{\boldsymbol{k}} c_{\boldsymbol{k}} = 1$ and $c_{\boldsymbol{k}} = 0$ for
$|\boldsymbol{k}| < 2$ yields
\begin{equation}
   \widetilde{S_{}}_J^{( 1 )} [ \{ \mu_p \} ] = - N' \ln G \Big( \sum_p \mu_p
   \Big),
\end{equation}
where $G ( x ) = \sum_{k \geqslant 2} x^k / k! = e^x - 1 - x$ is the
generating function of the ensemble. Reverting to original variables is easy.
Via the dual transform we obtain
\begin{equation}
   S_J^{( 1 )} [ N', M' ] = \min_{\{ \mu_p \}} \bigg\{ - M'  \sum_p \ln \mu_p
   + \tilde{S}_J^{( 1 )} [ \{ \mu_p \} ] \bigg\},
\end{equation}
and for $\ln N_J$ we obtain
\begin{equation}
   S_J [ N', M' ] = \min_{\{ \mu_p \}} \bigg\{ M' \sum_p \ln \frac{M'}{\mu_p}
   + N' \ln G \Big( \sum_p \mu_p \Big) \bigg\} - KM' .
\end{equation}
Clearly, the minimum is permutation-symmetric: $\mu_p = \mu / K$.
Equivalently,
\begin{equation}
   S_J [ N', M' ] = \min_{\mu} \left\{ KM' \ln \frac{KM'}{\mu} + N' \ln G (
   \mu ) \right\} - KM' .
\end{equation}
Comparison with (\ref{eq:KXOR-q}), (\ref{eq:KXOR-M}) gives $\mu = K \alpha q^{K - 1}$. Note that substituting
$G ( x ) = e^x$ (meaning no constraints on degrees of variables) gives $\mu =
KM' / N'$ and $S_J [ N', M' ] = KM' \ln N'$ as expected.

We now turn to computing the logarithm of $\mathcal{N}_{s, J}$. Binary
variables can take values of $0$ and $1$. Since these can be mapped onto $+ 1$
and $- 1$, with exclusive-or replaced by a product, from now
on we shall succinctly refer to values taken by variables as $+$ and $-$. For
each realization of disorder and variable assignment, we ascribe a type
$\boldsymbol{\sigma}$ to each clause according to the values of the variables inside
that clause; $\boldsymbol{\sigma}$ is a vector with $K$ elements with $\sigma_p
\in \{ +, - \}$. For the time being we fix the number of clauses of each type
$M_{\boldsymbol{\sigma}}'$ (remember that $M_{\boldsymbol{\sigma}} = 0$ unless
$\prod_p \sigma_p = +$ for the ferromagnetic model). In addition
to its value $s \in \{ +, - \}$, we ascribe to each variable a vector $\boldsymbol{k}$ of $K 2^K$
elements; $k_{\boldsymbol{\sigma}}^p$ denotes the number of clauses of type
$\boldsymbol{\sigma}$ in which that variable appears in $p$-th position. Having
fixed $M_{\boldsymbol{\sigma}}'$ and $N_{s,\boldsymbol{k}}'$ we discover that the
contribution to $\mathcal{N}_{s, J}$ is given by the product of $3$ factors:
\begin{enumerate}
  \item $N' ! \big/ \sum_{s,\boldsymbol{k}} N_{s,\boldsymbol{k}}' !$ for the number of
  ways to arrange the variables into classes.

  \item $\prod_{p,\boldsymbol{\sigma}} \left[ M_{\boldsymbol{\sigma}}' ! \Big/
  \prod_{s,\boldsymbol{k}} ( k_{\boldsymbol{\sigma}}^p ! )^{N_{s,\boldsymbol{k}}'}
  \right]$ for the number of ways to rearrange the variables within the clauses.

  \item $M' ! \big/ \prod_{\boldsymbol{\sigma}} M_{\boldsymbol{\sigma}}' !$ for the
  number of ways to assign types to clauses.
\end{enumerate}
The associated entropy
\begin{equation}
   S_{s, J} [ c_{s,\boldsymbol{k}} ] = - N' \sum_{s,\boldsymbol{k}}
   c_{s,\boldsymbol{k}} \ln \Big[ c_{s,\boldsymbol{k}}
   \prod_{p,\boldsymbol{\sigma}} k_{\boldsymbol{\sigma}}^p ! \Big] + K
   \sum_{\boldsymbol{\sigma}} \left( M_{\boldsymbol{\sigma}}' \ln
   M_{\boldsymbol{\sigma}}' - M_{\boldsymbol{\sigma}}' \right) + M' \ln M' -
   \sum_{\boldsymbol{\sigma}} M_{\boldsymbol{\sigma}}' \ln M_{\boldsymbol{\sigma}}'
\end{equation}
is to be maximized under the constraints
\begin{equation} \sum_{s,\boldsymbol{k}} k_{\boldsymbol{\sigma}}^p c_{s,\boldsymbol{k}} =
   M_{\boldsymbol{\sigma}}' \end{equation}
and the requirement that $c_{s,\boldsymbol{k}} = 0$ unless
$\sum_{p,\boldsymbol{\sigma}} k_{\boldsymbol{\sigma}}^p \geqslant 2$. Another
important constraint is that unless $\sigma_p = s$, we require
$k_{\boldsymbol{\sigma}}^p = 0$.

The optimization of the first part of the entropy is best accomplished through
the use of the dual transformation. The dual parameters are $\{ - \ln
\mu_{\boldsymbol{\sigma}}^p \}$:
\begin{eqnarray}
   \tilde{S}_{s, J}^{( 1 )} [ \{ \mu_{\boldsymbol{\sigma}}^p \} ] & = & \min_{\{
   M_{\boldsymbol{\sigma}}^p \}} \bigg\{ - \sum_{p,\boldsymbol{\sigma}}
   M_{\boldsymbol{\sigma}}^p \ln \mu_{\boldsymbol{\sigma}}^p \nonumber \\
   && - \max_{\{
   c_{s,\boldsymbol{k}} \}} \Big\{ - N' \sum_{s,\boldsymbol{k}}
   c_{s,\boldsymbol{k}} \ln \Big[ c_{s,\boldsymbol{k}}
   \prod_{p,\boldsymbol{\sigma}} k_{\boldsymbol{\sigma}}^p ! \Big] \Big|
   \sum_{\boldsymbol{k}} k_{\boldsymbol{\sigma}}^p c_{s,\boldsymbol{k}} =
   M_{\boldsymbol{\sigma}}^p \Big\} \bigg\} .
\end{eqnarray}
After simplifications we can rewrite
\begin{equation}
   \tilde{S}_{s, J}^{( 1 )} [ \{ \mu_{\boldsymbol{\sigma}}^p \} ] = N' \ln
   \bigg[ G \Big( \sum_{p,\boldsymbol{\sigma}} \frac{1 + \sigma_p}{2}
   \mu_{\boldsymbol{\sigma}}^p \Big) + G \Big( \sum_{p,\boldsymbol{\sigma}}
   \frac{1 - \sigma_p}{2} \mu_{\boldsymbol{\sigma}}^p \Big) \bigg] .
\end{equation}
The argument of the first $G$ is a sum restricted to $\sigma_p = +$, and the
argument of the second $G$ is a sum restricted $\sigma_p = -$. It is
convenient to introduce
\begin{equation}
   \mu_{\pm} = \sum_{p,\boldsymbol{\sigma}} \frac{1 \pm \sigma_p}{2}
   \mu_{\boldsymbol{\sigma}}^p .
\end{equation}
Reverting the dual transformation we can obtain
\begin{eqnarray}
  S_{s, J} [ N', \{ M_{\boldsymbol{\sigma}}' \} ] & = & \min_{\{
  \mu_{\boldsymbol{\sigma}}^p \}} \bigg\{ \sum_{p,\boldsymbol{\sigma}}
  M_{\boldsymbol{\sigma}}' \ln
  \frac{M_{\boldsymbol{\sigma}}'}{\mu_{\boldsymbol{\sigma}}^p} + N' \ln \bigg[ G
  \Big( \sum_{p,\boldsymbol{\sigma}} \frac{1 + \sigma_p}{2}
  \mu_{\boldsymbol{\sigma}}^p \Big) + G \Big( \sum_{p,\boldsymbol{\sigma}}
  \frac{1 - \sigma_p}{2} \mu_{\boldsymbol{\sigma}}^p \Big) \bigg] \bigg\} \nonumber \\
  &  & -KM' + M' \ln M' - \sum_{\boldsymbol{\sigma}} M_{\boldsymbol{\sigma}}' \ln
  M_{\boldsymbol{\sigma}}' .
\end{eqnarray}
It is convenient to define
\begin{equation}
   \mathcal{M}_{\pm} = \sum_{p,\boldsymbol{\sigma}} \frac{1 + \sigma_p}{2}
   M_{\boldsymbol{\sigma}}' .
\end{equation}
The expression for the entropy can be simplified to
\begin{eqnarray}
   S_{s, J} [ N', \{ M_{\boldsymbol{\sigma}}' \} ] & = & \min_{\mu_{\pm}} \left\{
   \mathcal{M}_+ \ln \frac{\mathcal{M}_+}{\mu_+} +\mathcal{M}_- \ln
   \frac{\mathcal{M}_-}{\mu_-} + \ln [ G ( \mu_+ ) + G ( \mu_- ) ] \right\} -
   KM' \nonumber \\
   && + M' \ln M' - \sum_{\boldsymbol{\sigma}} M_{\boldsymbol{\sigma}}' \ln
   M_{\boldsymbol{\sigma}}' .
\end{eqnarray}
The expression for the annealed entropy $S_{\textrm{ann}} = S_{s, J} - S_J$ thus
reads
\begin{eqnarray}
  S_{\textrm{ann}} [ N', \{ M_{\boldsymbol{\sigma}}' \} ] & = & \min_{\mu_{\pm}}
  \left\{ \mathcal{M}_+ \ln \frac{\mathcal{M}_+}{\mu_+} +\mathcal{M}_- \ln
  \frac{\mathcal{M}_-}{\mu_-} + \ln [ G ( \mu_+ ) + G ( \mu_- ) ] \right\} \nonumber \\
  &&-\min_{\mu} \left\{ KM' \ln \frac{KM'}{\mu} + \ln G ( \mu ) \right\} + M' \ln
  M' - \sum_{\boldsymbol{\sigma}} M_{\boldsymbol{\sigma}}' \ln
  M_{\boldsymbol{\sigma}}'.
\end{eqnarray}
This expression has to be maximized with respect to $M_{\boldsymbol{\sigma}}'$.
As a first step, we would like to maximize the third term $S_{\textrm{ann}}^{( 3
)} = M' \ln M' - \sum_{\boldsymbol{\sigma}} M_{\boldsymbol{\sigma}}' \ln
M_{\boldsymbol{\sigma}}'  $ keeping $M'$ and $\mathcal{M}_+
-\mathcal{M}_-$ fixed. Its dual is
\begin{equation}
   \tilde{S}_{\textrm{ann}}^{( 3 )} ( h ) = \min_{M_{\boldsymbol{\sigma}}'}
   \Big\{ - h (\mathcal{M}_+ -\mathcal{M}_- ) - M' \ln M' +
   \sum_{\boldsymbol{\sigma}} M_{\boldsymbol{\sigma}}' \ln M_{\boldsymbol{\sigma}}'
   \, \Big| \, M_{\boldsymbol{\sigma}}' = M' \Big\}
\end{equation}
Let $\epsilon_{\boldsymbol{\sigma}} \in \{ 0, 1 \}$ determine whether the clause
of type $\boldsymbol{\sigma}$ is permitted ($\epsilon_{\boldsymbol{\sigma}} = 1$)
or not ($\epsilon_{\sigma} = 0$). For the ferromagnetic model
$\epsilon_{\boldsymbol{\sigma}} = \frac{1 + \prod_p   \sigma_p}{2}$. For
$\tilde{S}_{\textrm{ann}}^{( 3 )}$ we obtain
\begin{equation}
   \tilde{S}_{\textrm{ann}}^{( 3 )} ( h ) = - M' \ln \sum_{\boldsymbol{\sigma}}
   \epsilon_{\boldsymbol{\sigma}} \mathrm{e}^{\left( \sum_p \sigma_p \right) h}
     = - M' \ln \frac{( 2 \cosh h )^K + ( 2 \sinh h )^K}{2},
\end{equation}
and the original $S_{\textrm{ann}}^{( 3 )} (\mathcal{M}_+,\mathcal{M}_- )$ is
given by
\begin{equation}
   S_{\textrm{ann}}^{( 3 )} [\mathcal{M}_+,\mathcal{M}_- ] = \min_h \Big\{ - h
   (\mathcal{M}_+ -\mathcal{M}_- ) + M' \ln \sum_{\boldsymbol{\sigma}}
   \epsilon_{\boldsymbol{\sigma}} \mathrm{e}^{\left( \sum_p \sigma_p
   \right) h} \Big\}.
\end{equation}
It is convenient to parameterize $\mathcal{M}_+$ and $\mathcal{M}_-$ by a
single parameter ($\mathcal{M}_+ +\mathcal{M}_- = KM'$ is a second
constraint). We can arbitrarily choose $h$ as such a parameter
\begin{eqnarray}
  \mathcal{M}_+ & = & M'  \frac{K + \mathrm{d} / \mathrm{d} h}{2} \ln
  \sum_{\boldsymbol{\sigma}} \epsilon_{\boldsymbol{\sigma}} \mathrm{e}^{\left( \sum_p
  \sigma_p   \right) h}  ,\\
  \mathcal{M}_- & = & M'  \frac{K - \mathrm{d} / \mathrm{d} h  }{2} \ln
  \sum_{\boldsymbol{\sigma}} \epsilon_{\boldsymbol{\sigma}} \mathrm{e}^{\left( \sum_p
  \sigma_p   \right) h}   .
\end{eqnarray}
For the case of the ferromagnetic model this becomes
\begin{equation}
   \mathcal{M}_{\pm} = KM' \mathrm{e}^{\pm h} \frac{\left( 2 \cosh h \right)^{K -
   1} \pm \left( 2 \sinh h \right)^{K - 1}}{( 2 \cosh h )^K + ( 2 \sinh h )^K}
   .
\end{equation}
Subsequently, we compute $S_{\textrm{ann}}$ as a function of $h$ and maximize
the expression with respect to $h$. For our special case we obtain that $h =
0$ gives the maximum to the expression  as long as $M' < N'$. For $h = 0$,
$S_{\textrm{ann}}$ takes a particularly simple form $S_{\textrm{ann}} = N' \ln 2 -
M' \ln 2$. Note that this is precisely the annealed entropy for $K$-XOR-SAT.
Therefore, the annealing approximation is correct up to $M' / N' = 1$, and the
corresponding connectivity of the original graph $\alpha \approx 0.918$ is
both an upper and a lower bound, i.e. the exact answer.

\section{\label{sec:KSAT}$K$-SAT model}

An instance of $K$-SAT is a set of $M$ clauses, each clause consisting of $K$
literals, where the literal is either one of $N$ variables $x_i$ or its
negation $\bar{x}_i$, each with probability $1 / 2$. The clause is satisfied
if at least one of the literals is $1$. Using boolean logic clause can be written
as an ``or'' of literals, e.g. $x_1 \vee \bar{x}_3 \vee \bar{x}_4$. A
formula is satisfied if all of its clauses are satisfied. For randomly
generated formulae, a satisfiability transition as a function of $M / N$ occurs
for some critical ratio $\alpha_c = M / N$. The exact location of this phase
transition is a major open problem.

A trivial upper bound is given by the annealing approximation. Notice that the
probability that a random clause is satisfied is independent of variable
assignment and equals $1 - 2^{- K}$. Correspondingly the annealed entropy
\begin{equation} \ln \mathbbm{E}[\mathcal{N}] = N \ln 2 + M \ln \left( 1 - 2^{- K} \right) .
\end{equation}
The annealing bound (where the annealed entropy is $0$) is hence $- 1 / \log_2
\left( 1 - 2^{- K} \right)$. For $K = 3$ this gives an upper bound of
$\alpha_u \approx 5.19$1, whereas numerical evidence places the transition at
$\alpha_c \approx 4.2$.

\subsection{Core for $K$-SAT problem}

Here the structure of disorder is more complex compared with the ferromagnetic
model since variables can appear both positively ($x$) and negatively
($\bar{x}$). To identify irrelevant clauses we use  the {\textit{pure literal}}
heuristic. Variables that appear only positively or only negatively can be set
to $1$ or $0$, respectively, to satisfy those clauses. Removing such ``pure''
literals together with clauses in which they appear for as long as possible
(as usual, we also remove variables that appear in no clauses) yields a much
smaller graph -- a core (see Fig. \ref{fig:trim3} below). Moreover, by the same logic, all cores with the same
number of variables $N'$ and clauses $M'$ and the condition that all variables
appear at least once positively and at least once negatively, are
equiprobable. We now turn to the subproblem of finding the expectation values of
$N'$ and $M'$ as a function of $\alpha = M / N$ that characterized the
original random formula.

\begin{figure}[!ht]
\includegraphics[width=2.5in]{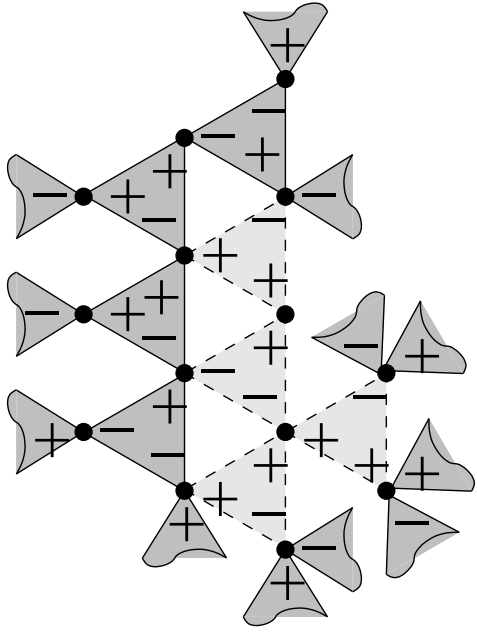}
\caption{\label{fig:trim3}Example of the trimming algorithm for 3-SAT. Variables are
represented graphically as vertices and clauses are represented as triangles. Signs ``+'' and
``-'' in triangles indicate whether the variable appears positively or negatively. Incomplete
triangles represent connections to the remainder of the graph (not shown). Lightly shaded clauses
are removed by the trimming algorithm.}
\end{figure}

As before, we use the notation $p = N' / N$ -- the probability that a randomly
chosen vertex belongs to the core. The set of variables in the core is denoted as
$\mathcal{C}$. We now introduce two different extensions of this set:
$\mathcal{C}'$ and $\bar{\mathcal{C}}'$ -- the minimal sets with following
properties
\begin{enumerate}
  \item $\mathcal{C} \subseteq \mathcal{C}'$ and $\mathcal{C} \subseteq
  \bar{\mathcal{C}}'$.

  \item If for some clause, $K - 1$ variables  have  a certain property, so should the
  remaining variable; the property being that the variable belongs to
  $\bar{\mathcal{C}}'$ if it appears positively or belongs to $\mathcal{C}'$ if
  it appears negatively.
\end{enumerate}
We also reserve the notation $q = |\mathcal{C}' |$ and $\bar{q} = |
\bar{\mathcal{C}}' |$. Also observe that $\mathcal{C}=\mathcal{C}' \cap
\bar{\mathcal{C}}'$.

Fix a variable $x_0$. It appears in $k$ clauses positively (as $x_0$) and in
$\bar{k}$ clauses negatively (as $\bar{x}_0$). The numbers $k$, $\bar{k}$ are
independent random variables distributed according to a Poisson distribution
with parameter $K \alpha / 2$. We assume that $q$ and $\bar{q}$ for the full
formula $\mathcal{F}$ are not different from $q'$ and $\bar{q}'$ for the
formula $\mathcal{F}'$ with variable $x_0$ deleted. Dropping primes we can
write self-consistency equations for $q$, $\bar{q}$:
\begin{eqnarray}
  q & = & \sum_{k = 0}^{\infty} \sum_{\bar{k} = 1}^{\infty} \mathrm{e}^{- K \alpha
  \left( \frac{q + \bar{q}}{2} \right)^{K - 1}} \frac{\left( \frac{K}{2}
  \alpha \frac{q + \bar{q}}{2} \right)^k}{k!}  \frac{\left( \frac{K}{2} \alpha
  \frac{q + \bar{q}}{2} \right)^{\bar{k}}}{\bar{k} !} = 1 - \mathrm{e}^{- \frac{K
  \alpha}{2} \left( \frac{q + \bar{q}}{2} \right)^{K - 1}}\\
  \bar{q} & = & \sum_{k = 1}^{\infty} \sum_{\bar{k} = 0}^{\infty} \mathrm{e}^{- K
  \alpha \left( \frac{q + \bar{q}}{2} \right)^{K - 1}} \frac{\left(
  \frac{K}{2} \alpha \frac{q + \bar{q}}{2} \right)^k}{k!}  \frac{\left(
  \frac{K}{2} \alpha \frac{q + \bar{q}}{2} \right)^{\bar{k}}}{\bar{k} !} = 1 -
  \mathrm{e}^{- \frac{K \alpha}{2} \left( \frac{q + \bar{q}}{2} \right)^{K - 1}}
\end{eqnarray}
Obviously $q = \bar{q}$ and a simpler equation could be written
\begin{equation} q = 1 - \mathrm{e}^{- \frac{K \alpha}{2} q^{K - 1}} . \end{equation}
Notice that this is identical to (\ref{eq:KXOR-q}) with the replacement $\alpha \rightarrow
\alpha / 2$. As a consequence, the core appears at exactly twice the threshold
for $K$-XOR-SAT (for $3$-XOR-SAT the core appears at $\alpha \approx 0.818$, and
for $3$-SAT it appears at $\alpha \approx 1.636$. This threshold was obtained earlier
(by a different method) in one of the first papers on lower bounds
for the satisfiability transition in $3$-SAT.)

To find $p$, the sums have to be restricted to $k \geqslant 1$ and $\bar{k}
\geqslant 1$ thus giving $p = q \bar{q}$. Hence
\begin{equation}
N' / N = q^2
\end{equation}
To find $M' / N$ we need to count the average degree of the variable in the
core
\begin{eqnarray}
   M' / N & = & \frac{1}{K} \sum_{k = 1}^{\infty} \sum_{\bar{k} = 1}^{\infty}
   \left( k + \bar{k} \right) \mathrm{e}^{- K \alpha \left( \frac{q + \bar{q}}{2}
   \right)^{K - 1}} \frac{\left( \frac{K}{2} \alpha \frac{q + \bar{q}}{2}
   \right)^k}{k!}  \frac{\left( \frac{K}{2} \alpha \frac{q + \bar{q}}{2}
   \right)^{\bar{k}}}{\bar{k} !} \nonumber \\
   & = & \alpha \Big( \frac{q + \bar{q}}{2}
   \Big)^{K - 1} \left( 1 - \mathrm{e}^{- \frac{K}{2} \alpha \left( \frac{q +
   \bar{q}}{2} \right)^{K - 1}} \right) .
\end{eqnarray}
Simplified, this becomes $M' / N = \alpha q^K$.

\subsection{Improved bound for $K$-SAT}

Now that the remaining clauses are correlated, the annealed entropy for $K$-SAT is
not as easily computed as for $K$-XOR-SAT. The technique parallels one used to
find the annealed entropy for the ferromagnetic model. We need to find the
logarithm of the number of disorders $\mathcal{N}_J$ and the logarithm of the
number of spin-disorder combinations $\mathcal{N}_{s, J}$. In contrast to
$K$-XOR-SAT, clauses acquire a type $\boldsymbol{\tau}$ -- a vector, elements of
which determine whether the variable in $p$-th position appears inverted or
not ($\tau_p \in \{ +, - \}$). Correspondingly, a vertex degree is now a
vector $\boldsymbol{k}$ with elements $k_{\boldsymbol{\tau}}^p$ describing the
number of appearances of a certain variable in the $p$-th position in clauses of
type $\boldsymbol{\tau}$. We fix the number of variables with given
$\boldsymbol{k}$: $N_{\boldsymbol{k}}'$ (corresponding fractions are
$c_{\boldsymbol{k}} = N_{\boldsymbol{k}}' / N'$). The number of disorders for
fixed $\left\{ N_{\boldsymbol{k}}' \right\}$ and $\left\{ M_{\boldsymbol{\tau}}'
\right\}$ is composed of the following factors:
\begin{enumerate}
  \item $N' ! \big/ \prod_{\boldsymbol{k}} N_{\boldsymbol{k}}' !$ for the number of
  ways to divide the variables into classes.

  \item $\prod_{p,\boldsymbol{\tau}} \left[ M_{\boldsymbol{\tau}}' ! \Big/
  \prod_{\boldsymbol{k}} ( k_{\boldsymbol{\tau}}^p ! )^{N_{\boldsymbol{k}}'}
  \right]$ for the number of ways to rearrange the variables among clauses.

  \item $M' ! \big/ \prod_{\boldsymbol{\tau}} M_{\boldsymbol{\tau}}' !
   $ for the number of permutations of clauses of various types.
\end{enumerate}
Taking the logarithm, we obtain
\begin{equation}
   S_J [ \{ c_{\boldsymbol{k}} \} ] = - N' \sum_{\boldsymbol{k}} c_{\boldsymbol{k}}
   \ln \Big[ c_{\boldsymbol{k}}  \prod_{p,\boldsymbol{\tau}}
   k_{\boldsymbol{\tau}}^p ! \Big] + K \sum_{\boldsymbol{\tau}}   \left(
   M_{\boldsymbol{\tau}}' \ln M_{\boldsymbol{\tau}}' - M_{\boldsymbol{\tau}}'
   \right) + M' \ln M' - \sum_{\boldsymbol{\tau}} M_{\boldsymbol{\tau}}' \ln
   M_{\boldsymbol{\tau}}' .
\end{equation}
We must optimize over $c_{\boldsymbol{k} \bar{\boldsymbol{k}}}$ taking into
account the constraint that $c_{\boldsymbol{k}} = 0$ if either $|\boldsymbol{k}| =
0$ or $| \bar{\boldsymbol{k}} | = 0$. We also have constraints $N'
\sum_{\boldsymbol{k}}   k_{\boldsymbol{\tau}}^p   c_{\boldsymbol{k}} =
M_{\boldsymbol{\tau}}'$. Introducing dual variables and a generating function $G
( x, \bar{x} ) = \left( e^x - 1 \right) \left( e^{\bar{x}} - 1 \right)$ we can
write
\begin{equation}
   S [ N', \{ M_{\boldsymbol{\tau}}' \} ] = \min_{\left\{
   \mu_{\boldsymbol{\tau}}^p \right\}} \bigg\{ \sum_{p,\boldsymbol{\tau}}
   M_{\boldsymbol{\tau}}' \ln
   \frac{M_{\boldsymbol{\tau}}'}{\mu_{\boldsymbol{\tau}}^p} + N' \ln G ( \mu,
   \bar{\mu} ) \bigg\} + M' \ln M' - \sum_{\boldsymbol{\tau}}
   M_{\boldsymbol{\tau}}' \ln M_{\boldsymbol{\tau}}' - KM',
\end{equation}
where
\begin{eqnarray}
  \mu & = & \sum_{p,\boldsymbol{\tau}} \frac{1 + \tau_p}{2}
  \mu_{\boldsymbol{\tau}}^p  ,\\
  \bar{\mu} & = & \sum_{p,\boldsymbol{\tau}} \frac{1 - \tau_p}{2}
  \mu_{\boldsymbol{\tau}}^p   .
\end{eqnarray}
Also introducing the quantities
\begin{eqnarray}
  \mathcal{M} & = & \sum_{p,\boldsymbol{\tau}} \frac{1 + \tau_p}{2}
  M_{\boldsymbol{\tau}}',\\
  \bar{\mathcal{M}} & = & \sum_{p,\boldsymbol{\tau}} \frac{1 - \tau_p}{2}
  M_{\boldsymbol{\tau}}',
\end{eqnarray}
we rewrite $S_J$ as
\begin{equation}
   S_J [ N', \{ M_{\boldsymbol{\tau}}' \} ] = \min_{\mu, \bar{\mu}} \left\{
   \mathcal{M} \ln \frac{\mathcal{M}}{\mu} + \bar{\mathcal{M}} \ln
   \frac{\bar{\mathcal{M}}}{\bar{\mu}} + N' \ln G ( \mu, \bar{\mu} ) \right\}
   + M' \ln M' - \sum_{\boldsymbol{\tau}} M_{\boldsymbol{\tau}}' \ln
   M_{\boldsymbol{\tau}}' .
\end{equation}
For convenience we will write $G ( \mu, \bar{\mu} ) = G_1 ( \mu ) G_1 (
\bar{\mu} )$. where $G_1 ( x ) = e^x - 1$. One can verify that $S_J$ is
maximized when $M_{\boldsymbol{\tau}}' = M' / 2^K$ and $\mu = \bar{\mu} =
\frac{K \alpha}{2} q^{K - 1}$.

Now we need to evaluate $\mathcal{N}_{s, J}$. This time the clauses are
parameterized by $\boldsymbol{\tau}$ -- the appearance of literals in a clause
as well as $\boldsymbol{\sigma}$ -- the particular assignments of variables. We
fix $\left\{ M_{\boldsymbol{\sigma} \boldsymbol{\tau}}' \right\}$ as well as
$\left\{ N_{\boldsymbol{k}}' \right\}$, with $\boldsymbol{k}$ being a vector with
$K 2^{2 K}$ elements: $k_{\boldsymbol{\sigma} \boldsymbol{\tau}}^p$ is the number
of appearances of a variable in clauses of type
$(\boldsymbol{\sigma},\boldsymbol{\tau})$ in the $p$-th position. The number
$\mathcal{N}_{s, J}$ can be broken into the following factors
\begin{enumerate}
  \item $N! \big/ \prod_{\boldsymbol{k}} N_{\boldsymbol{k}}' !$

  \item $\prod_{p,\boldsymbol{\sigma},\boldsymbol{\tau}} \left[
  M_{\boldsymbol{\sigma} \boldsymbol{\tau}}' ! \Big/ \prod_{\boldsymbol{k}} \left(
  k_{\boldsymbol{\sigma} \boldsymbol{\tau}}^p ! \right)^{N_{\boldsymbol{k}}'}
  \right]$

  \item $M' ! \big/ \prod_{\boldsymbol{\sigma},\boldsymbol{\tau}} M_{\boldsymbol{\sigma}
  \boldsymbol{\tau}}' !$
\end{enumerate}
Enforcing constraints $N' \sum_{s,\boldsymbol{k}} k_{\boldsymbol{\sigma}
\boldsymbol{\tau}}^p c_{s,\boldsymbol{k}}   = M_{\boldsymbol{\sigma}
\boldsymbol{\tau}}'$ as well as constraints on the vector $\boldsymbol{k}$, i.e. that
$\sum_{p,\boldsymbol{\sigma},\boldsymbol{\tau}} \frac{1 \pm \tau_p}{2}
k_{\boldsymbol{\sigma} \boldsymbol{\tau}}^p \geqslant 1$ and that
$c_{s,\boldsymbol{k}} = 0$ if for some $p,\boldsymbol{\sigma},\boldsymbol{\tau}$
$k_{\boldsymbol{\sigma} \boldsymbol{\tau}}^p > 0$ and $\sigma_p \neq s$, we are
able to cast the expression for $S_{s, J} [ N', \{ M_{\boldsymbol{\sigma}
\boldsymbol{\tau}}' \} ]$ in a simple form
\begin{eqnarray}
   S_{s, J} [ N', \{ M_{\boldsymbol{\sigma} \boldsymbol{\tau}}' \} ] & = &
   \min_{\mu_{\pm}, \bar{\mu}_{\pm}} \Big\{
   \sum_{p,\boldsymbol{\sigma},\boldsymbol{\tau}} M_{\boldsymbol{\sigma}
   \boldsymbol{\tau}}' \ln \frac{M_{\boldsymbol{\sigma}
   \boldsymbol{\tau}}'}{\mu_{\boldsymbol{\sigma} \boldsymbol{\tau}}^p} + \ln [ G (
   \mu_+, \bar{\mu}_+ ) + G ( \mu_-, \bar{\mu}_- ) ] \Big\} \nonumber \\
   && + M' \ln M' -
   \sum_{\boldsymbol{\sigma},\boldsymbol{\tau}} M_{\boldsymbol{\sigma}
   \boldsymbol{\tau}}' \ln M_{\boldsymbol{\sigma} \boldsymbol{\tau}}' - KM',
\end{eqnarray}
where
\begin{eqnarray}
  \mu_{\pm} & = & \sum_{p,\boldsymbol{\sigma},\boldsymbol{\tau}} \frac{1 +
  \tau_p}{2}  \frac{1 \pm \sigma_p}{2} \mu_{\boldsymbol{\sigma}
  \boldsymbol{\tau}}^p  ,\\
  \bar{\mu}_{\pm} & = & \sum_{p,\boldsymbol{\sigma},\boldsymbol{\tau}} \frac{1 -
  \tau_p}{2}  \frac{1 \pm \sigma_p}{2} \mu_{\boldsymbol{\sigma}
  \boldsymbol{\tau}}^p .
\end{eqnarray}
Also introducing
\begin{eqnarray}
  \mathcal{M}_{\pm} & = & \sum_{p,\boldsymbol{\sigma},\boldsymbol{\tau}} \frac{1 +
  \tau_p}{2}  \frac{1 \pm \sigma_p}{2} M_{\boldsymbol{\sigma} \boldsymbol{\tau}}',
   \\
  \bar{\mathcal{M}}_{\pm} & = & \sum_{p,\boldsymbol{\sigma},\boldsymbol{\tau}}
  \frac{1 - \tau_p}{2}  \frac{1 \pm \sigma_p}{2} M_{\boldsymbol{\sigma}
  \boldsymbol{\tau}}',
\end{eqnarray}
we can rewrite the first part of $S_{s, J}$ as
\begin{eqnarray}
   S_{s, J}^{( 1 )} [ N', \{ M_{\boldsymbol{\sigma} \boldsymbol{\tau}}' \} ] =
   \min_{\mu_{\pm}, \bar{\mu}_{\pm}} \Big\{ && \mathcal{M}_+ \ln
   \frac{\mathcal{M}_+}{\mu_+} +\mathcal{M}_- \ln \frac{\mathcal{M}_-}{\mu_-}
   + \bar{\mathcal{M}}_+ \ln \frac{\bar{\mathcal{M}}_+}{\bar{\mu}_+} +
   \bar{\mathcal{M}}_- \ln \frac{\bar{\mathcal{M}}_-}{\bar{\mu}_-} \nonumber \\
   && + N' \ln [ G (\mu_+, \bar{\mu}_+ ) + G ( \mu_-, \bar{\mu}_- ) ] \Big\} \label{eq:K-SsJ1}
\end{eqnarray}
Next, we optimize the expression $M' \ln M' -
\sum_{\boldsymbol{\sigma},\boldsymbol{\tau}} M_{\boldsymbol{\sigma}
\boldsymbol{\tau}}' \ln M_{\boldsymbol{\sigma} \boldsymbol{\tau}}'$ subject to fixed
$\mathcal{M}_+$, $\mathcal{M}_-$, $\bar{\mathcal{M}}_+$, $\bar{\mathcal{M}}_-$.
Introducing dual variables $- h, - h', - h''$ coupled to $\mathcal{M}_+
-\mathcal{M}_- - \bar{\mathcal{M}}_+ + \bar{\mathcal{M}}_-$, $\mathcal{M}_+
-\mathcal{M}_- + \bar{\mathcal{M}}_+ - \bar{\mathcal{M}}_-$ and $\mathcal{M}_+
+\mathcal{M}_- - \bar{\mathcal{M}}_+ - \bar{\mathcal{M}}_-$ respectively, the
optimized expression becomes
\begin{eqnarray}
   S_{s, J}^{( 2 )} [ N', \mathbf{\mathcal{M}} ] && =
   \min_{h, h', h''} \Big\{ - h ( \mathcal{M}_+ -\mathcal{M}_- -
   \bar{\mathcal{M}}_+ + \bar{\mathcal{M}}_- ) - h' ( \mathcal{M}_+
   -\mathcal{M}_- + \bar{\mathcal{M}}_+ - \bar{\mathcal{M}}_- ) \nonumber \\
   &&- h''
   ( \mathcal{M}_+ +\mathcal{M}_- - \bar{\mathcal{M}}_+ -
   \bar{\mathcal{M}}_- ) - M' \ln \sum_{\boldsymbol{\sigma}
   \boldsymbol{\tau}} \epsilon_{\boldsymbol{\sigma} \boldsymbol{\tau}}
   \mathrm{e}^{\left( \sum_p \sigma_p \tau_p \right) h + \left( \sum_p \sigma_p
   \right) h' + \left( \sum_p \tau_p \right) h''}   \Big\} \nonumber \\
   &&
\end{eqnarray}
where $\epsilon_{\boldsymbol{\sigma} \boldsymbol{\tau}} \in \{ 0, 1 \}$ determines
whether the clause is permitted. For the case of $K$-SAT we only prohibit
combinations $\prod_p \frac{1 + \sigma_p \tau_p}{2} = 1$. We can
express $\mathcal{M}_+,\mathcal{M}_-, \bar{\mathcal{M}}_+,
\bar{\mathcal{M}}_-$ in terms of $h, h'$ and $h''$ and substitute into (\ref{eq:K-SsJ1}).
Consequently, maximization over $h, h', h''$ will be performed. It can be shown
that the maximum necessarily corresponds to $h' = h'' = 0$ leading to further
simplifications:
\begin{equation}
   \sum_{\boldsymbol{\sigma},\boldsymbol{\tau}} \epsilon_{\boldsymbol{\sigma}
   \boldsymbol{\tau}} \mathrm{e}^{\left( \sum_p \sigma_p \tau_p \right) h} = ( 2
   \cosh h )^K - \mathrm{e}^{Kh},
\end{equation}
and we can show that $\bar{\mathcal{M}}_+ =\mathcal{M}_-$ and
$\bar{\mathcal{M}}_- =\mathcal{M}_+$. (As a result $\bar{\mu}_+ = \mu_-$ and
$\bar{\mu}_- = \mu_+$).
\begin{eqnarray}
   S_{\textrm{ann}} = \max_h \Big\{ && + \min_{\mu} \big\{
   2\mathcal{M}_+ \ln \frac{2\mathcal{M}_+}{\mu} + N' \ln G_1 ( \mu ) \big\}
   + \min_{\mu} \big\{ 2\mathcal{M}_- \ln \frac{2\mathcal{M}_-}{\mu} + N' \ln
   G_1 ( \mu ) \big\} \nonumber \\
   && - 2 \min_{\mu} \left\{ \frac{KM'}{2} \ln \frac{KM' /
   2}{\mu} + N' \ln G_1 ( \mu ) \right\} + N'\ln 2 - KM' \ln 2 \nonumber \\
   && - h ( 2\mathcal{M}_+ -
   2\mathcal{M}_- ) + M' \ln \left[ ( 2 \cosh h )^K - \mathrm{e}^{Kh} \right]
   \Big\},
\end{eqnarray}
where $\mathcal{M}_{\pm}$ are the functions of $h$:
\begin{equation}
   2\mathcal{M}_{\pm} = \frac{1}{2} KM' \pm \frac{1}{2}  \frac{\mathrm{d}}{\mathrm{d}
   h} \ln \sum_{\boldsymbol{\sigma} \boldsymbol{\tau}} \epsilon_{\boldsymbol{\sigma}
   \boldsymbol{\tau}} \mathrm{e}^{\left( \sum_p \sigma_p \tau_p \right) h},
\end{equation}
or, substituting $\epsilon_{\boldsymbol{\sigma} \boldsymbol{\tau}}$ for $K$-SAT
\begin{eqnarray}
  2\mathcal{M}_+ & = & KM' \mathrm{e}^h  \frac{( 2 \cosh h )^{K - 1} - \mathrm{e}^{( K
  - 1 ) h}}{( 2 \cosh h )^K - \mathrm{e}^{Kh}},\\
  2\mathcal{M}_- & = & KM' \mathrm{e}^{- h}  \frac{( 2 \cosh h )^{K - 1} }{( 2
  \cosh h )^K - \mathrm{e}^{Kh}} .
\end{eqnarray}
We also verify that the maximum of the complete expression corresponds to $h =
0$. As a result

\begin{eqnarray}
   S_{\textrm{ann}} & = & \min_{\mu}
   \left\{ 2\mathcal{M}_+ \ln \frac{2\mathcal{M}_+}{\mu} + N' \ln G_1 ( \mu )
   \right\} + \min_{\mu} \left\{ 2\mathcal{M}_- \ln \frac{2\mathcal{M}_-}{\mu}
   + N' \ln G_1 ( \mu ) \right\} \nonumber \\
   && - 2 \min_{\mu} \left\{ \frac{KM'}{2} \ln
   \frac{KM' / 2}{\mu} + N' \ln G_1 ( \mu ) \right\} + N' \ln 2 +M' \ln \left[ 1 - 2^{- K} \right]
\end{eqnarray}
Solving $S_{\textrm{ann}} = 0$ translates into an upper bound for $K = 3$ of
$\alpha_u' \approx 5.189$ -- a rather insignificant improvement over
straightforward annealing approximation.

\section{\label{sec:1inKSAT}Positive $1$-in-$K$-SAT model}

In this model, we have a set of clauses, each clause involving $K$ variables
that can take values $0$ or $1$. A clause is satisfied if the sum of values of
variables is exactly $1$.  A formula is satisfied if all the clauses that constitute
it are satisfied. A related problem was considered in \cite{farhi} in the context
of the quantum adiabatic algorithm, which served as the main motivation for present analysis.
For a randomly generated formula, the satisfiability
transition occurs for some critical clause-to-variable ratio $\alpha = M / N$.
The easiest upper bound is obtained using the straightforward annealing
approximation. For the logarithm of the expected number of solutions we obtain
\begin{equation}
   \ln \mathbbm{E}[\mathcal{N}] = \max_m \bigg\{ - N \Big( \frac{1 + m}{2}
   \ln \frac{1 + m}{2} + \frac{1 - m}{2} \ln \frac{1 - m}{2} \Big) + M \ln
   \Big[ 3 \frac{1 - m}{2} \left( \frac{1 + m}{2} \right)^2 \Big] \bigg\},
\end{equation}
where we identified $m = 1$ with having all variables assigned a value of $0$,
$m = - 1$ with having all variables set to $1$, and intermediate values of $m$
being the appropriate mixture.

The annealed entropy becomes $0$ at the critical threshold $\alpha_u \approx
0.805$. We now seek to improve upon this simplistic approximation.

\subsection{Core for positive $1$-in-$K$-SAT}

\begin{figure}[!ht]
\includegraphics[width=6in]{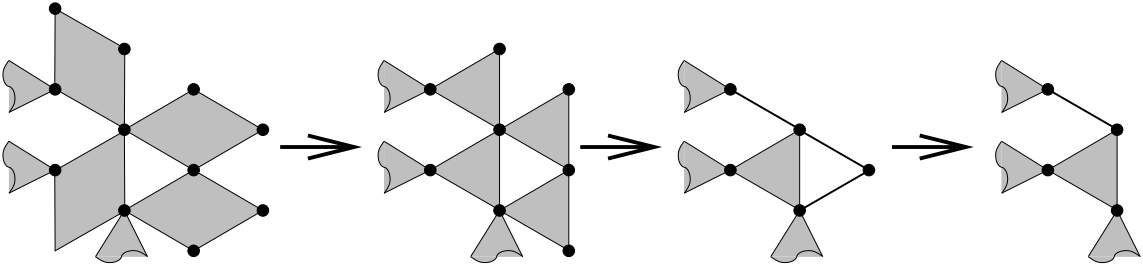}
\caption{\label{fig:trim2}Example of the trimming algorithm for 1-in-4-SAT. Variables are represented
graphically as vertices and 4-, 3- and 2-clauses are represented as rhombi, triangles
and edges correspondingly. Incomplete polygons represent connections to the remainder of the graph.
The figure depicts evolution of part of the graph under the trimming algorithm.}
\end{figure}

The structure of the core for positive $1$-in-$K$-SAT is more complex than
what we have seen before. As before variables of degree $0$ are eliminated.
Similarly variables of degree $1$ are removed, although we are no longer
justified in removing the clause in which variable appears. Instead, the
corresponding $K$-clause has to be replaced with a $( K - 1 )$-clause. The
latter is deemed to be satisfied if the sum of variables in it is either $0$
or $1$. Then the remaining variable could always be set to either $0$ or $1$
so that the sum of all $K$ variables is exactly $1$. Similarly, if any
variable has degree $1$ and appears in a $( K - 1 )$-clause, the latter can be
converted to a $( K - 2 )$-clause and so on. For all clauses of length less than
$K$, the criterion for satisfiability is that the sum of variables be either
$0$ or $1$. Finally, we identify variables that appear only in $2$ clauses.
Setting any such variable to $0$ will satisfy all $2$-clauses. Thus, such
variables and clauses in which they appear can be eliminated. This process
continues until we are left with a subformula where the degree of each variable
 is $\geqslant 2$ and no variable appears only in $2$-clauses (see Fig. \ref{fig:trim2}).

For any
fixed $N'$ and a set of $\{ M_k' \}$ (with $k = 2, \ldots, K$) -- the number
of clauses of length $k$ -- all subformulae that satisfy aforementioned
constraints are equally probable. The values $N' / N$ and $\{ M_k' / N \}$ are
self-averaging and their means will be computed shortly.

As before, we introduce the following notation. $\mathcal{C}$ denotes the set of
variables that belong to the core. In addition to $\mathcal{C}$ we introduce
sets $\mathcal{C}_2'$ and $\mathcal{C}'$. The sets shall have the following
properties:
\begin{enumerate}
  \item $\mathcal{C} \subseteq \mathcal{C}' \subseteq \mathcal{C}_2'$.

  \item If $2$ variables in some clause belong to $\mathcal{C}'$, then all
  variables in that clause belong to $\mathcal{C}'$.

  \item If $1$ variable in some clause belongs to $\mathcal{C}_2'$, then all
  variables in that clause belong to $\mathcal{C}_2'$.
\end{enumerate}
We reserve the notation $p = |\mathcal{C}| / N$, $q = |\mathcal{C}' | / N$ and
$q_2 = |\mathcal{C}_2' | / N$. As before, we single out a single variable
$x_0$ and study the probability that the variable belongs to classes
$\mathcal{C}$, $\mathcal{C}'$ or $\mathcal{C}_2'$. The number of clauses in
which the variable appears is Poisson with parameter $K \alpha$. The variable
$x_0$ is in $\mathcal{C}'$ if for at least one clause in which $x_0$ appears
at least two variables among the $K - 1$ remaining variables belong to
$\mathcal{C}_2'$.
\begin{equation}
   q = 1 - \exp \left[ - K \alpha \left( 1 - ( 1 - q_2 )^{K - 1} - ( K - 1 )
   q_2 ( 1 - q_2 )^{K - 2} \right) \right],
\end{equation}
where we have used the fact that the probability that among randomly chosen $K
- 1$ variables the probability that at least two belong to $\mathcal{C}_2'$ is
$1 - ( 1 - q_2 )^{K - 1} - ( K - 1 ) q_2 ( 1 - q_2 )^{K - 2}$.

The variable $x_0$ is in $\mathcal{C}_2'$ if for at least one clause, at least
one variable among the other $( K - 1 )$ variables belongs to $\mathcal{C}'$ or at
least two variables belong to $\mathcal{C}_2'$. The probability of that is $1
- ( 1 - q_2 )^{K - 1} - ( K - 1 ) ( q_2 - q ) ( 1 - q )^{K - 2}$. The second
self-consistency equation is thus
\begin{equation}
   q_2 = 1 - \exp \left[ - K \alpha \left( 1 - ( 1 - q_2 )^{K - 1} - ( K - 1 )
   ( q_2 - q ) ( 1 - q_2 )^{K - 2} \right) \right] .
\end{equation}
Consider clauses in which the variable $x_0$ appears. Let us call those clauses in
which at least two variables appear in $\mathcal{C}_2'$ type-$1$ clauses, and
those clauses in which one variable belongs to $\mathcal{C}'$ -- type-$2$
clauses. Variable $x_0$ is in $\mathcal{C}$ if it appears in two or more
type-$1$ or type-$2$ clauses, and at least one type-$1$ clause. Therefore, we
should have
\begin{equation} p = 1 - \mathrm{e}^{- K \alpha p_1} - K \alpha p_1 \mathrm{e}^{- K \alpha p_2}, \end{equation}
where
\begin{eqnarray}
  p_1 & = & 1 - ( 1 - q_2 )^{K - 1} - ( K - 1 ) q_2 ( 1 - q_2 )^{K - 2},\\
  p_2 & = & 1 - ( 1 - q_2 )^{K - 1} - ( K - 1 ) ( q_2 - q ) ( 1 - q_2 )^{K -
  2} .
\end{eqnarray}
To find the number of $k$-clauses in the core $M_k'$, compute the average
$k$-degree of variable $x_0$, i.e. the number of $k$-clauses in which it
appears. We readily obtain the following formulae:
\begin{eqnarray}
  M_2' / N & = & \binom{K}{2} \alpha q^2 ( 1 - q_2 )^{K - 2},\\
  M_k' / N & = & \binom{K}{k} \alpha q_2^k ( 1 - q_2 )^{K - k}, \textrm{for } k
  \geqslant 3.
\end{eqnarray}

\subsection{Improved bound for positive $1$-in-$K$-SAT}

As before, we compute $\mathcal{N}_J$ -- the number of disorders, subject to
fixed $N'$ and $\{ M_k' \}$, under the condition that each variable has a
degree of at least two, and that no variable appears in $2$-clauses
exclusively. Introduce a vector of length $K - 1$ of vertex degrees $( k_2,
\ldots, k_K )$, with elements being the number of $k$-clauses in which the
variable appears. We prohibit vertices with$\sum_{i = 3}^K k_i = 0$ or
$\sum_{i = 2}^K k_i = 1$. The corresponding generating function
\begin{equation}
   G (\boldsymbol{x}) = \sum_{\{ k_i \}}' \prod_{i = 2}^K \frac{( x_i
   )^{k_i}}{k_i !} = \mathrm{e}^{\sum_{i = 2}^K x_i} - \mathrm{e}^{x_2} - \sum_{i =
   3}^K x_i .
\end{equation}
It is convenient to write $G (\boldsymbol{x}) = G_2 \left( x_2, \sum_{k = 3}^K
x_k \right)$, where $G_2 ( x, y ) = \mathrm{e}^{x + y} - e^x - y$.

We proceed to counting the number of disorders with fixed $N'$ and $\{ M_k'
\}$. It is convenient to introduce the quantities $N_{\boldsymbol{k}_2 \ldots
\boldsymbol{k}_K}'$ that count the number of vertices; indices $k_i^p$ being the
number of appearances in $p$-th position in a clause of length $i$. Starting
from
\begin{equation}
   S_J [ c_{\boldsymbol{k}_2 \ldots \boldsymbol{k}_K} ] = - N'
   \sum_{\boldsymbol{k}_2 \ldots \boldsymbol{k}_K} c_{\boldsymbol{k}_2 \ldots
   \boldsymbol{k}_K} \ln \Big[ c_{\boldsymbol{k}_2 \ldots \boldsymbol{k}_K}
   \prod_{i, p} k_i^p ! \Big] + \sum_{k = 2}^K k ( M_k' \ln M_k' - M_k' )
\end{equation}
and optimizing over $c_{\boldsymbol{k}_2 \ldots \boldsymbol{k}_K}$ subject to
constraints on degrees as well as the set of constraints
\begin{equation}
   N' \sum_{\boldsymbol{k}_2 \ldots \boldsymbol{k}_K} k_i^p c_{\boldsymbol{k}_2
   \ldots \boldsymbol{k}_K} = M_i'
\end{equation}
we obtain
\begin{equation}
   S_J [ N', \{ M_k' \} ] = \min_{\{ \mu_k \}} \Big\{ \sum_{k = 2}^K kM_k'
   \ln \frac{kM_k'}{\mu_k} + N' \ln G ( \{ \mu_k \} ) \Big\} - \sum_{k =
   2}^K kM_k'
\end{equation}
Using the relation $G ( \{ \mu_k \} ) = G_2 \big( \mu_2, \sum_{k = 3}^K \mu_k
\big)$ rewrite
\begin{eqnarray}
   S_J [ N', \{ M_k' \} ] & = & \min_{\mu_2, \mu} \bigg\{ 2 M_2' \ln \frac{2
   M_2'}{\mu_2} + \Big( \sum^K_{k = 3} kM_k' \Big) \ln \frac{\sum_{k
   = 3}^K kM_k' }{\mu} + N' \ln G_2 ( \mu_2, \mu ) \bigg\} \nonumber \\
   && - \sum_{k = 2}^K kM_k' .
\end{eqnarray}
Now, compute the total number of disorders and variable assignments compatible
with them. Now the clauses of each length have to be subdivided into types
$\boldsymbol{\sigma}_2$ through $\boldsymbol{\sigma}_K$, according to the variable
assignments in the corresponding clause. We arrange variables into classes
according to their value $s \in \{ +, - \}$ and a vector $(\boldsymbol{k}_2,
\ldots,\boldsymbol{k}_K )$, with $k_{\boldsymbol{\sigma}_i}^p$ being the number of
appearances of a variable in a clause of length $i$ and type
$\boldsymbol{\sigma}_i$ in $p$-th position. The number $\mathcal{N}_{s, J}$ is
given as a product of three factors
\begin{enumerate}
  \item $N' ! \big/ \prod_{s,\boldsymbol{k}_2 \ldots \boldsymbol{k}_K}
  N_{s,\boldsymbol{k}_2 \ldots \boldsymbol{k}_K}' !$ for the number of ways to
  rearrange the variables into classes

  \item $\prod_{i = 2}^K \prod_p \prod_{\boldsymbol{\sigma}_i} \left[
  M_{\boldsymbol{\sigma}_i}' ! \Big/ \prod_{s,\boldsymbol{k}_2 \ldots \boldsymbol{k}_K}
  ( k_{\boldsymbol{\sigma}_i}^p ! )^{N_{s,\boldsymbol{k}_2 \ldots \boldsymbol{k}_K}}
  \right]$ for the number of ways to rearrange variables inside the clauses.

  \item $\prod_{i = 2}^K \left[ M_i' ! \big/ \sum_{\boldsymbol{\sigma}_i}
  M_{\boldsymbol{\sigma}_i}' ! \right]$ for the number of ways to rearrange
  clauses.
\end{enumerate}
For the entropy we obtain
\begin{eqnarray}
   S_{s,J} [ c_{s,\boldsymbol{k}_2 \ldots \boldsymbol{k}_K} ] & = & - N'
   \sum_{s,\boldsymbol{k}_2 \ldots \boldsymbol{k}_K} c_{s,\boldsymbol{k}_2 \ldots
   \boldsymbol{k}_K} \ln \Big[ c_{s,\boldsymbol{k}_2 \ldots \boldsymbol{k}_K}
   \prod_{i,\boldsymbol{\sigma}_{\boldsymbol{i}}, p} k_{\boldsymbol{\sigma}_i}^p !
   \Big] + \sum_{i,\boldsymbol{\sigma}_i} i \left( M_{\boldsymbol{\sigma}_i}'
   \ln M_{\boldsymbol{\sigma}_i}' - M_{\boldsymbol{\sigma}_i}' \right) \nonumber \\
   && + \sum_i M_i' \ln M_i' - \sum_{i,\boldsymbol{\sigma}_i}
   M_{\boldsymbol{\sigma}_i}' \ln M_{\boldsymbol{\sigma}_i}' .
\end{eqnarray}
We must note the constraints
\begin{equation}
   N' \sum_{\boldsymbol{k}_2 \ldots \boldsymbol{k}_K} k_{\boldsymbol{\sigma}_i}^p =
   M_{\boldsymbol{\sigma}_i}^p
\end{equation}
as well as constraints on variable value ($k_{\boldsymbol{\sigma}_i}^p \neq 0
\Rightarrow \sigma_i^p = s$) and on the degrees of the variables ($\sum_{i = 3}^K
|\boldsymbol{k}_i | \geqslant 1$ and $\sum_{i = 2}^K |\boldsymbol{k}_i | \geqslant
2$). With the aid of the generating function and the dual variables we can write
\begin{eqnarray}
   S_{s, J} [ N', \{ M_k \} ] & = & \min_{\{ \mu_{\boldsymbol{\sigma}_k}^p \}}
   \bigg\{ \sum_{k,\boldsymbol{\sigma}_k, p} M_{\boldsymbol{\sigma}_k}' \ln
   \frac{M_{\boldsymbol{\sigma}_k}'}{\mu_{\boldsymbol{\sigma}_k}^p} + N' \ln
   \left[ G \left( \left\{ \mu_{k +} \right\} \right) + G \left( \left\{
   \mu_{k -} \right\} \right) \right] \bigg\} \nonumber \\
   && + \sum_k M_k' \ln M_k' -
   \sum_{k,\boldsymbol{\sigma}_k} M_{\boldsymbol{\sigma}_k}' \ln
   M_{\boldsymbol{\sigma}_k}' - \sum_k kM_k',
\end{eqnarray}
where we have written $\mu_{k \pm} = \sum_{p,\boldsymbol{\sigma}_k} \frac{1 +
\sigma_k^p}{2} \mu_{\boldsymbol{\sigma}_k}^p$. Also, introducing
$\mathcal{M}_{k \pm} = \sum_{p,\boldsymbol{\sigma}_k} \frac{1 + \sigma_k^p}{2}
M_{\boldsymbol{\sigma}_k}'$ the first subexpression can be simplified
to
\begin{equation}
   S_{s, J}^{( 1 )} [ N', \mathbf{\mathcal{M}} ] = \min_{\left\{ \mu_k^{\pm} \right\}}
   \bigg\{ \sum_k \Big( \mathcal{M}_{k +} \ln \frac{\mathcal{M}_{k +}}{\mu_{k
   +}} +\mathcal{M}_{k -} \ln \frac{\mathcal{M}_{k -}}{\mu_{k -}} \Big) + N'
   \ln \left[ G \left( \left\{ \mu_{k +} \right\} \right) + G \left( \left\{
   \mu_{k -} \right\} \right) \right] \bigg\},
\end{equation}
and using $G_2$ can be rewritten as
\begin{eqnarray}
   S_{s, J}^{( 1 )} [ N', \mathbf{\mathcal{M}} ] = && \min_{\mu_{2 \pm}, \mu_{\pm}} \bigg\{
   \mathcal{M}_{2 +} \ln \frac{\mathcal{M}_{2 +}}{\mu_{2 +}} +\mathcal{M}_{2
   +} \ln \frac{\mathcal{M}_{2 -}}{\mu_{2 -}} + \Big( \sum_{k = 3}^K
   \mathcal{M}_{k +} \Big) \ln \frac{\sum_{k = 3}^K \mathcal{M}_{k
   +}}{\mu_+} \nonumber \\
   && + \Big( \sum_{k = 3}^K \mathcal{M}_{k -} \Big) \ln
   \frac{\sum_{k = 3}^K \mathcal{M}_{k -}}{\mu_-} + \ln\left[ G_2 \left(
   \mu_{2 +}, \mu_- \right) + G_2 \left( \mu_{2 -}, \mu_- \right) \right]
   \bigg\}
\end{eqnarray}
In correspondence with the different treatment afforded to $2$-clauses and
$k$-clauses for $k \geqslant 3$, we introduce two fields $- h_2$ and $- h$
coupled to $\mathcal{M}_{2 +} -\mathcal{M}_{2 -}$ and $\sum_{k = 3}^K \left(
\mathcal{M}_{k +} -\mathcal{M}_{k -} \right)$ correspondingly. The dual of
second part of $S_{s, J}$ is
\begin{eqnarray}
   \tilde{S}_{s, J}^{( 2 )} [ h_2, h ] = \min_{\left\{
   M_{\boldsymbol{\sigma}_k}' \right\}} \bigg\{ && - h_2 \left( \mathcal{M}_{2 +}
   -\mathcal{M}_{2 -} \right) - h \sum_{k = 3}^K \left( \mathcal{M}_{k +}
   -\mathcal{M}_{k -} \right) \nonumber \\
   && - \sum_{k = 2}^K \Big( M_k' \ln M_k' -
   \sum_{\boldsymbol{\sigma}_k} M_{\boldsymbol{\sigma}_k}' \ln
   M_{\boldsymbol{\sigma}_k}' \Big) \bigg\} .
\end{eqnarray}
Note that for $k = K$ only $\sum_p \sigma_K^p = K - 1$ is allowed, while for
$k < K$, $\sum_p \sigma_k^p = K$ and $\sum_p \sigma_k^p = K - 1$ are both
allowed. After proper minimizations we obtain
\begin{equation}
   \tilde{S}_{s, J}^{( 2 )} [ h_2, h ] = M_2' \ln \left( 2 + \mathrm{e}^{h_2}
   \right) + \sum_{k = 3}^{K - 1} M_k' \ln \left( k \mathrm{e}^{( k - 1 ) h} +
   \mathrm{e}^{kh} \right) + M_K' \ln \left( K \mathrm{e}^{Kh} \right) .
\end{equation}
We can express $\left\{ M_k^{\pm} \right\}$ in terms of $h_2$ and $h$ via
\begin{eqnarray}
  \mathcal{M}_{2 \pm} & = & M_2' \pm \frac{1}{2}  \frac{\partial}{\partial
  h_2}  \tilde{S}_{s, J}^{( 2 )} [ h_2, h ],\\
  \sum_{k = 3}^K \mathcal{M}_{k \pm} & = & \sum_{k = 3}^K kM_k' \pm
  \frac{1}{2}  \frac{\partial}{\partial h}  \tilde{S}_{s, J}^{( 2 )} [ h_2, h
  ] .
\end{eqnarray}
The entire expression for the annealed entropy is then written as
\begin{eqnarray}
   \tilde{S}_{\textrm{ann}} [ h_2, h ] & = & \min_{\mu_{2 \pm}, \mu_{\pm}}
   \Bigg\{ \mathcal{M}_{2 +} \ln \frac{\mathcal{M}_{2 +}}{\mu_{2 +}}
   +\mathcal{M}_{2 +} \ln \frac{\mathcal{M}_{2 -}}{\mu_{2 -}} + \Big( \sum_{k
   = 3}^K \mathcal{M}_{k +} \Big) \ln \frac{\sum_{k = 3}^K \mathcal{M}_{k
   +}}{\mu_+} \nonumber \\
   && + \Big( \sum_{k = 3}^K \mathcal{M}_{k -} \Big) \ln
   \frac{\sum_{k = 3}^K \mathcal{M}_{k -}}{\mu_-} + \ln \left[ G_2 \left(
   \mu_{2 +}, \mu_- \right) + G_2 \left( \mu_{2 -}, \mu_- \right) \right]
   \Bigg\} \nonumber \\
   && - \min_{\mu_2, \mu} \bigg\{ 2 M_2' \ln \frac{2 M_2'}{\mu_2} +
   \Big( \sum^K_{k = 3} kM_k' \Big) \ln \frac{\sum_{k = 3}^K kM_k'
   }{\mu} + N' \ln G_2 ( \mu_2, \mu ) \bigg\} \nonumber \\
   && - h_2 (\mathcal{M}_{2 +}
   -\mathcal{M}_{2 -} ) - h \sum_{k = 3}^K (\mathcal{M}_{k +}
   -\mathcal{M}_{k -} ) \nonumber \\
   && + M_2' \ln ( 2 + \mathrm{e}^{h_2} ) + \sum_{k = 3}^{K - 1}
   M_k' \ln \left( k \mathrm{e}^{( k - 1 ) h} + \mathrm{e}^{kh} \right) + M_K' \ln
   \left( K \mathrm{e}^{Kh} \right)
\end{eqnarray}
Maximization over $h_2, h$ and solving $S_{\textrm{ann}} = 0$ gives an upper
bound for the satisfiability transition. For $K = 3$ we obtain $\alpha_u \approx
0.644$. This compares favorably to $\alpha_c \approx 0.625$ observed in
simulations and beats the previous best upper bound of $\alpha_u \approx 0.727$
\cite{moore}.

\section{\label{sec:Simulations}Simulation results}

In this section we present experimental results on random positive 1-in-3-SAT
instances.  Using the Davis-Putnam (DP) algorithm (see Appendix \ref{app:A})
we study the crossover point and the computation complexity.  We also identify
experimentally the position of the phase transition.

\subsection{The Crossover Point}
\label{sec:crossover}

\begin{figure}[!ht]
\includegraphics[width=6in]{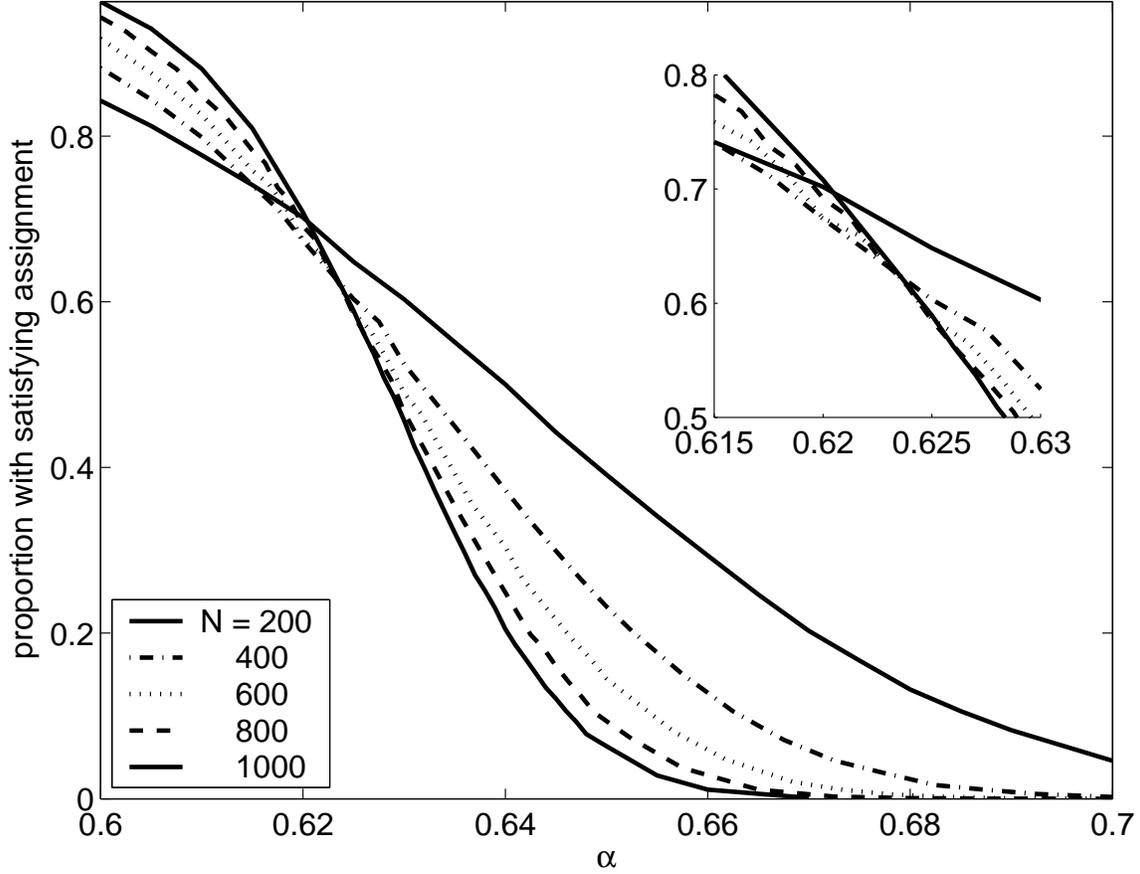}
\caption{Proportion of problem instances with a satisfying assignment}
\label{fig:crossover}
\end{figure}

The major feature of a phase transition in a satisfiability problem is
the presence of a threshold in $\alpha$, below which almost all random
problem instances are solvable, and above which almost no random
problem instances are.  Figure \ref{fig:crossover}  shows a plot of
the proportion of random problem instances that have a satisfying
assignment, versus $\alpha$, for various values of $N$.  The
proportions are based on running the DP algorithm on 50,000 random
problem instances for each value of $N$ and $\alpha$.  The expected
features are present.  The sharpness of the phase transition increases
with $N$, and the point at which the curve crosses the line where the
proportion of instances with a satisfying assignment equals $0.5$
decreases with $N$.

\begin{figure}[!ht]
\includegraphics[width=6in]{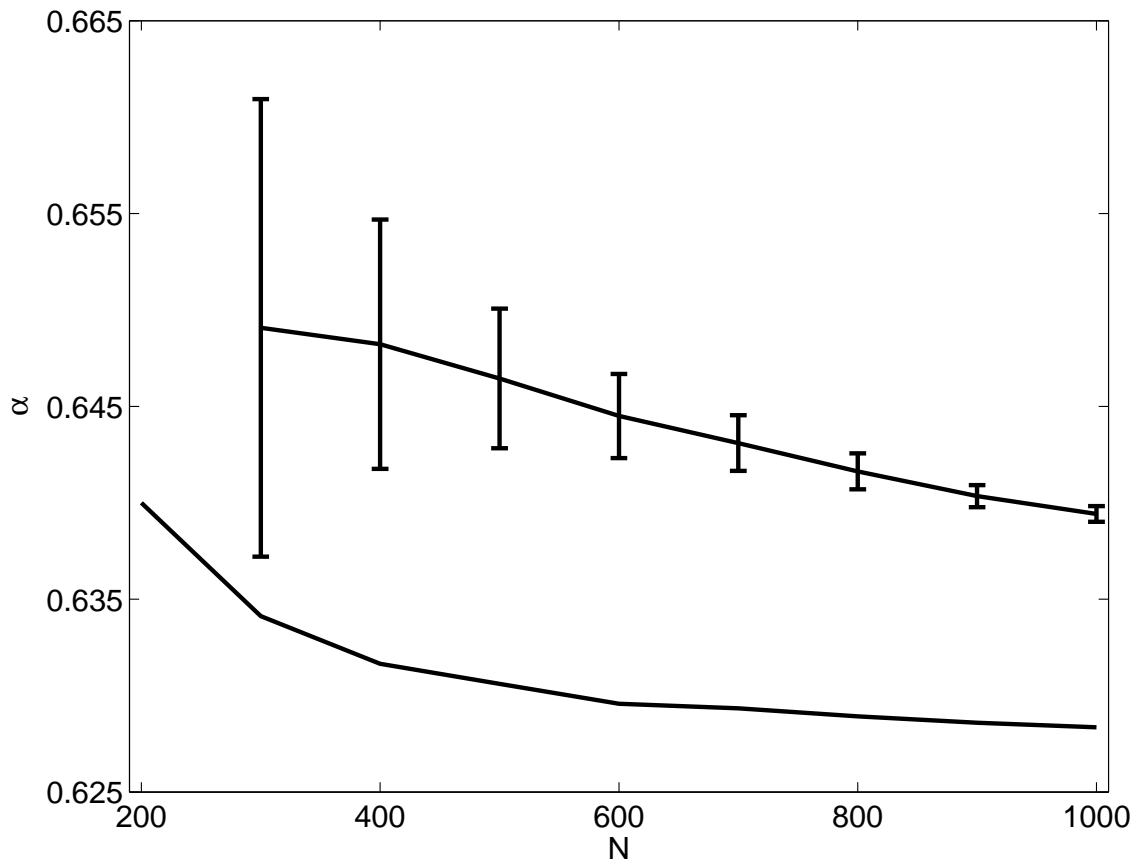}
\caption{Top curve: plot of the maximum complexity of the DP algorithm.
Lower curve: the position of the crossing point of proportion with
satisfying assignment = 0.5}
\label{fig:phase_transition}
\end{figure}

Experimentally the crossover point is at $\alpha_c\approx 0.625$,
slightly lower than the upper bound of $\alpha_u \approx 0.644$ computed in
section \ref{sec:1inKSAT}. In figure \ref{fig:phase_transition}
(lower curve) we plot the value of $\alpha$ for which 50\% of the
problem instances were satisfiable as a function of the number of
bits.  The curve appears to have an asymptote around $\alpha_c\approx 0.625$.

\subsection{Complexity of the Davis-Putnam Algorithm}
\label{sec:dp_complexity}

\begin{figure}[!ht]
\includegraphics[width=6in]{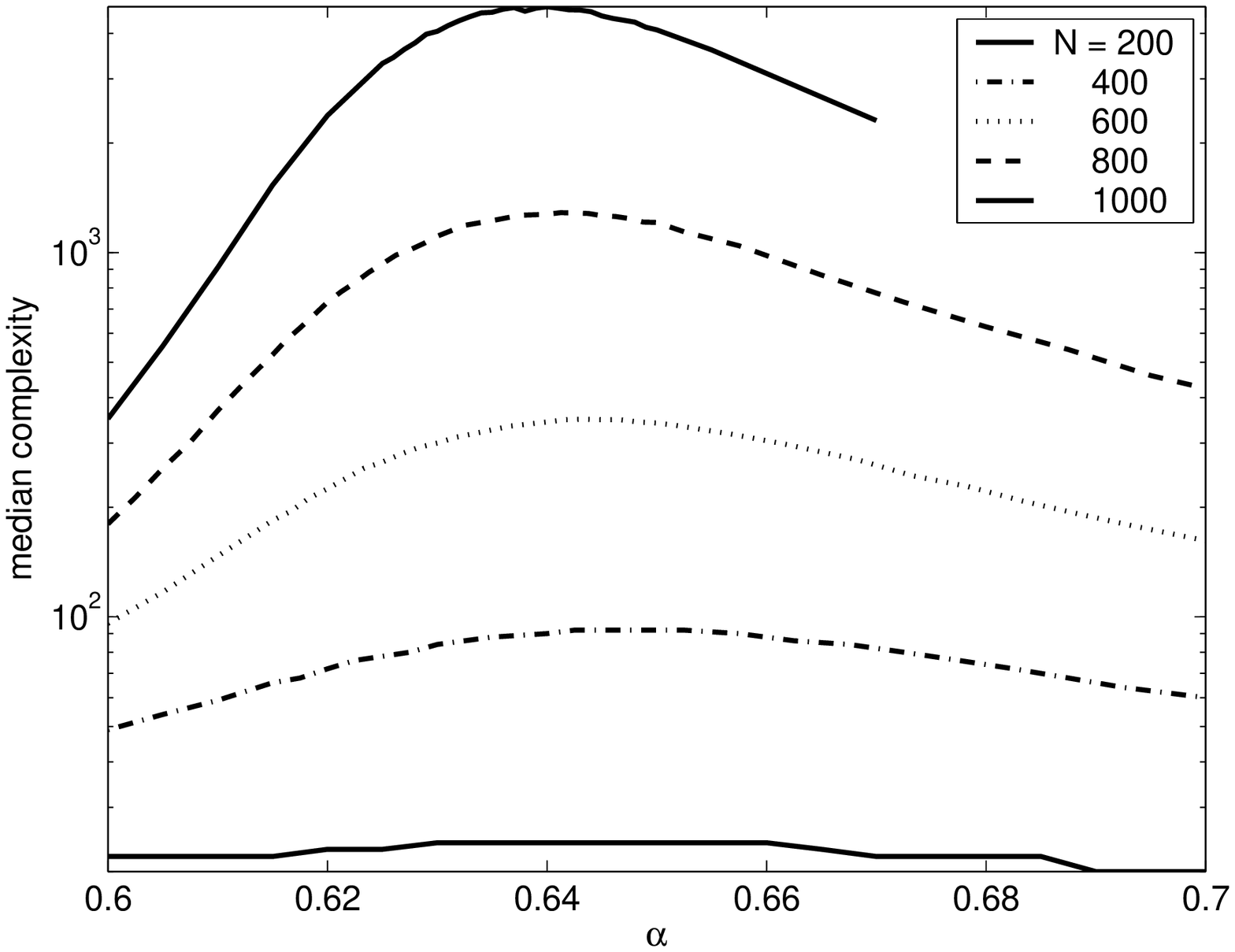}
\caption{Computational complexity of DP}
\label{fig:dp_complexity_1}
\end{figure}

Figure \ref{fig:dp_complexity_1} shows plots of the median complexity
of the Davis-Putnam (DP) algorithm (complexity is defined as the
number of calls to the function \verb!Find_Model! displayed in Table \ref{tab:DP}).
The median was taken over 50,000 random problem instances.  As expected, because
the DP algorithm is complete, its performance scales exponentially
with problem size, $N$. Note also that the value of $\alpha$ for which
the maximum complexity occurs is above $\alpha_c$, and slowly reduces
as $N$ increases.  In figure \ref{fig:phase_transition} (upper curve) we
plot the position of the maximum complexity and its uncertainty.  we
note that for the range of values of $N$ considered, it does not appear
to have converged to an asymptote, but the curve does not appear to
contradict our earlier result of $\alpha_c\approx 0.625$.

Fitting an exponential law to the peak complexity gives
$C=6.13\exp(0.0067\times N)$, a very slow rate of increase -- an order of
magnitude slower than reported results on the complexity of DP applied
to 3-SAT \cite{ksat2}.

\section{\label{sec:Summary}Summary}

In this paper we have proposed a new method for analyzing subgraphs (subformulae) of
the random graph (formula) subject to simple geometric constraints.
For every constraint satisfaction problem one can identify a core -- a subformula
that is satisfiable if and only if the original formula was satisfiable.
In fact simplifying the original formula is typically a first step before applying
general-purpose algorithms such as the Davis-Putnam routine or simulated annealing,
and the best algorithms use it. This may become an essential tool for the analysis of
``smart'' algorithms that perform transformations on the instance of the problem
or even on intermediate steps. We have also applied the methods used in the present
paper for the approximate analysis of the quantum adiabatic algorithm for positive
$1$-in-$K$-SAT problem \cite{self}.

We have also tried to estimate the satisfiability transition from the above for three problems:
$K$-XOR-SAT, $K$-SAT and Positive $1$-in-$K$-SAT. The results for $K=3$ are as follows:
$\alpha_u \approx 0.918$ for $K$-XOR-SAT (exact), $\alpha_u \approx 5.189$ for $K$-SAT
(vs. $\alpha_c \approx 4.2$ experimentally) and $\alpha_u \approx 0.644$ for
positive $1$-in-$K$-SAT (vs. $\alpha_c \approx 0.625$ experimentally).

The bound for $K$-SAT was an insignificant
improvement over the annealing approximation despite deleting irrelevant clauses that
contribute to the entropy. Results for $K$-XOR-SAT and $1$-in-$K$-SAT were quite good.
Note that random $1$-in-$K$-SAT (where variables may appear in clauses either positively
or negatively with probability $1/2$, akin to $K$-SAT) is quite simple. The satisfiability
transition coincides with percolation, and algorithms solve the problem very efficiently
in the satisfiable phase. A precise way to state this is that the dynamical transition coincides
with the satisfiability transition, shrinking the difficult region. This is not the case for
positive $1$-in-$K$-SAT that we consider, where most likely $\alpha_d<\alpha_c$.

That the annealing approximation for the simplified formula fails to predict the correct transition
suggests that a large number of solutions remains up to the satisfiability threshold. In all likelihood
these individual solutions are well-separated, which may explain the poor performance of algorithms.
We conjecture that random instances of positive $1$-in-$K$-SAT are significantly simpler
to solve than those of $K$-SAT. This view is partly supported by simulations. Also observe
that the answer for $K$-XOR-SAT -- a polynomial problem -- is exact.

\section{Acknowledgments}
This work was supported in part by the National Security Agency
(NSA) and Advanced Research and Development Activity (ARDA) under
Army Research Office (ARO) contract number
ARDA-QC-P004-J132-Y03/LPS-FY2003, we also want to acknowledge the
partial support of NASA CICT/IS program.

\appendix

\section{\label{app:A}The Davis-Putnam Algorithm}

The Davis-Putnam (DP) algorithm \cite{davis60}, or a variation, is regarded
as the most efficient {\em complete} algorithm for satisfiability
problems.  An outline of the DP algorithm is given in table
\ref{tab:DP} \cite{ksat2}.  The version we used varies from
this outline in one major respect.  We perform a sort of the variables
before the first call to {\verb!Find_Model!}, sorting on the number of clauses
which use the variable.  This was found to produce, on average, a very
large speed-up in the algorithm's execution.

The {\verb!unit_propagate!} step of the algorithm is also extremely efficient
for the $1$-in-$K$-SAT problem.  Once one variable in a clause is set to
$1$, the value of the other two variables is fixed, and extensive
propagation often occurs.  Also, because a single variable in a clause
being set to $1$ determines the other two variables in the clause, we
call {\verb!Find_Model( theory AND x )!} first.

\begin{table}[!ht]
\caption{Outline of the Davis-Putnam Algorithm}
\label{tab:DP}
\begin{verbatim}
Find_Model( theory )
  unit_propagate( theory );
  if contradiction discovered return(false);
  else if all variables are valued return(true);
  else {
    x = some unvalued variable;
    return( Find_Model( theory AND x ) OR
            Find_Model( theory AND NOT x ) );
  }
\end{verbatim}
\end{table}

\end{document}